\documentclass[aps,superscriptaddress,twocolumn]{revtex4-1}

\usepackage{amsmath,amsfonts,amsthm,amsbsy}

\usepackage{graphicx}
\usepackage{braket}

\usepackage[svgnames,x11names]{xcolor}

\usepackage{tikz}
\usetikzlibrary{arrows}
\usetikzlibrary{intersections}

\definecolor{ct_red}{HTML}{E32636}
\definecolor{ct_green}{rgb}{0.47,0.67,0.19}
\definecolor{ct_blue}{rgb}{0,0.45,0.74}

\definecolor{ct2_green}{HTML}{9FF781}
\definecolor{ct2_green_dark}{HTML}{088A08}

\definecolor{blue_M}{rgb}{0.37,0.51,0.71}
\definecolor{orange_M}{rgb}{0.88,0.61,0.14}

\usepackage[pdftex,colorlinks=true,linkcolor=blue,citecolor=blue,urlcolor=blue]{hyperref}

\newcommand{\ee}{{\mathrm e}}
\newcommand{\ii}{{\mathrm i}}
\newcommand{\dd}{{\mathrm d}}

\begin{document}

\title{Anomalous bulk-edge correspondence in continuous media}

\author{C. Tauber}
\email{tauberc@phys.ethz.ch}
\affiliation{Institute for Theoretical Physics, ETH Z¨urich, Wolfgang-Pauli-Str. 27, CH-8093 Z¨urich}

\author{P. Delplace}
\author{A. Venaille}
\affiliation{Univ Lyon, Ens de Lyon, Univ Claude Bernard, CNRS, Laboratoire de Physique, F-69342
Lyon
}

\date{\today}

\begin{abstract}

Topology plays an increasing role in physics beyond the realm of topological insulators in condensed mater. From geophysical fluids to active matter, acoustics or photonics, a growing family of systems presents topologically protected chiral edge modes. The number of such modes should coincide with the bulk topological invariant (e.g. Chern number) defined for a sample without boundary, in agreement with the bulk-edge correspondence. However this is not always the case when dealing with continuous media where there is no small scale cut-off. The number of edge modes actually depends on the boundary condition, even when the bulk is properly regularized, showing an apparent paradox where the bulk-edge correspondence is violated. In this paper we solve this paradox by showing that the anomaly is due to {ghost} edge modes hidden in the asymptotic part of the spectrum. We provide a general formalism based on scattering theory to detect all edge modes properly, so that the bulk-edge correspondence is restored. We illustrate this approach through the odd-viscous shallow-water model and the massive Dirac Hamiltonian, and discuss the physical consequences.
\end{abstract}

\pacs{Valid PACS appear here}

\maketitle

\section{Introduction}

Bulk-edge correspondence is a hallmark of topology in physics. When there exists a topological number associated to an infinite and gaped system (the bulk), it states that topologically protected edge modes appear in a sample with a boundary, and \textit{vice versa}. These modes are confined near the boundary, robust to many perturbations and their number coincide with the bulk topological quantity. 

The relevance of topology in physics starts with the Quantum Hall Effect, where it was realised that both bulk and edge picture were associated to topological quantities \cite{thouless1982quantized,laughlin1981quantized,halperin1982quantized}, that actually coincide \cite{hatsugai1993chern}. It was then widely expanded through the field of topological insulators \cite{hasan2010colloquium}, where bulk-edge correspondence was studied and proved in systems with various dimensions and symmetries \cite{hatsugai2009bulk,isaev2011bulk,graf2013bulk,avila2013topological}, in presence of (strong) disorder \cite{schulz2000simultaneous,elgart2005equality,prodan2016bulk,Graf2018}, or for periodically driven (Floquet) systems \cite{rudner2013anomalous,asboth2014chiral,graf2018bulk,shapiro2018strongly}.

In the context of condensed matter the bulk-edge correspondence usually focuses on lattice models thank to the tight-binding approximation. However this problem was somehow overlooked in continuous models, namely beyond this approximation or when there is no underlying lattice structure. Apart from continuous electronic models, e.g. the Landau Hamiltonian, topology has also appeared in virtually all fields of physics, from superfluids \cite{volovik1988analogue} to photonics \cite{raghu2008analogs,rechtsman2013photonic,lu2014topological,peano2015topological,silveirinha2018proof} or molecular spectra \cite{faure2000topological}, among others. These ideas have then been applied to the realm of classical fluid and solid mechanics,  including elasticity \cite{prodan2009topological,kane2014topological,susstrunk2015observation},  acoustics \cite{yang2015topological,fleury2016floquet,peri2018axial}, geophysical and astrophysical flows \cite{delplace2017topological,perrot2018topological},  plasma \cite{jin2016topological,gao2016photonic,jin2018magnetically,silveirinha2016bulk}, or active matter \cite{shankar2017topological,souslov2017topological,souslov2018topological}. There, a continuous medium description is natural.

One example is the two-dimensional shallow-water model describing Earth atmospheric and oceanic layers \cite{delplace2017topological,tauber2018odd}, {and its formal analogs encountered in} active matter and plasma {physics} \cite{souslov2018topological}, as well as in optical systems \cite{van2018photon}. It appears as a paradigmatic (spin 1) three band model, by analogy with the celebrated (spin 1/2) Dirac Hamiltonian \cite{volovik1988analogue}. In the context of geophysical fluids, the topology of the shallow-water model was recently revealed. Due to the sign change of Coriolis force, the existence of uni-directional waves propagating near the equator could be interpreted as topologically protected \cite{delplace2017topological}. More recently it was shown that a topological (Chern) number can be assigned to the bulk problem for this flow, up to a regularization by an odd-viscous term \cite{souslov2018topological,tauber2018odd}. Indeed, in contrast to condensed matter where quasi-momenta live on a compact torus (Brillouin Zone), the momentum {(or the wave number)} is usually unbounded in continuous models in the absence of any cut-off and has to be properly regularized. {In this way, as in condensed matter, a meaningful bulk topological number can be defined that is expected to rule the bulk-boundary correspondence in continuous media and thus predict the number of chiral edge modes. }

\begin{figure}[htbp]
\centering
\includegraphics[scale=0.81]{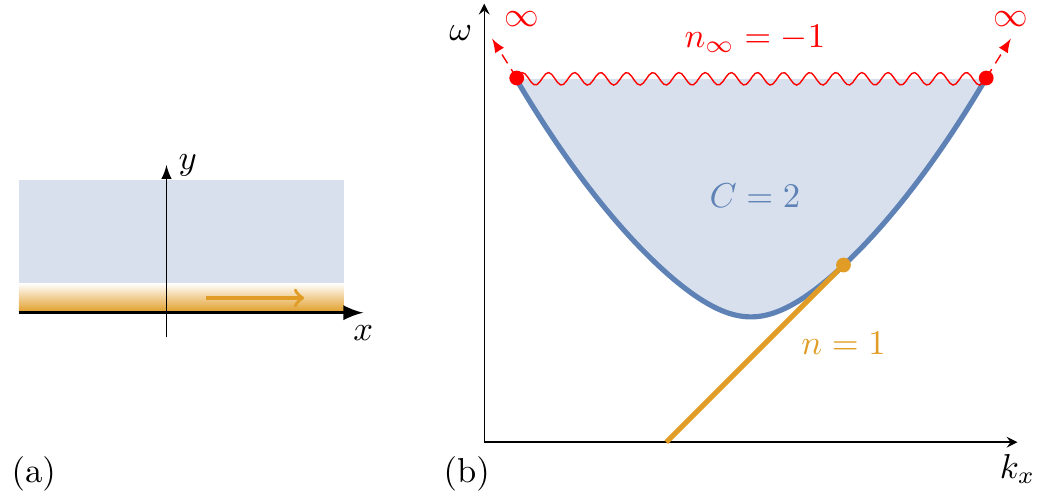}
\caption{(a) Continuous model with a sharp boundary. (b) Typical spectrum: the delocalized bulk modes form a band (in blue) that is gaped below but unbounded above. It has a topological (Chern) number $C$. The gaped region may host $n$ edge modes (in orange) that are confined near the boundary and uni-directional. Although topological, this number apparently depends on the boundary condition. However the bulk-edge correspondence $C=n-n_\infty$ is always satisfied if we take into account possible {ghost} modes at infinity (in red). \label{fig:intro} }
\end{figure}

However, as we shall see, the regularization of the bulk does not implies the same for the edge problem. Indeed, in {the shallow water model,} we observe that the number of edge modes depends on the boundary condition, be it with odd-viscous terms \cite{tauber2018odd}, or without it \cite{iga1995transition}. This looks suspicious compared to the expected topological nature of these modes in the presence of odd viscosity, and raises the apparent paradox of a violation of the bulk-edge correspondence. This anomaly is not restricted to the shallow-water model and was actually already noticed in other two-dimensional continuous models, e.g. in the valley Quantum Hall effect \cite{li2010marginality} or compressible stratified fluids \cite{iga2001transition}, {that are both effectively well described by a Dirac Hamiltonian}.

In this paper we propose a solution to this paradox and restore the bulk-edge correspondence for continuous models with a sharp boundary. The crucial observation is that in such models, neither the longitudinal momentum nor the frequency {(or energy)} are bounded, so that the usual way to count the edge modes might miss the asymptotic area of the spectrum, see Figure \ref{fig:intro}. Thus we provide an alternative formalism based on scattering theory, that counts properly the usual edge modes but also allows to detect  {\textit{ghost edge modes}}  that could be hidden at {infinite frequencies in the spectrum}. Applying it to several boundary conditions, we show that this is indeed the case so that the bulk-edge correspondence is  {restored} when all the modes, {including the ghost modes that are not visible in the spectrum at finite frequency and momentum,} are properly taken into account, {thus  revealing an \textit{anomalous bulk-boundary correspondence} for continuous media.} Note that this approach works beyond the illustrative choice of the shallow-water model and applies similarly to any continuous model as long as the bulk is properly regularized, such as the compactified Dirac Hamiltonian that we also tackle at the end.

Scattering theory has  been previously involved into the definition of topological quantities in tight-binding discrete models, through two independent ways. The first way was to probe the presence of edge modes of a topological sample through scattering from \emph{outside} the sample \cite{meidan2011topological,fulga2011scatt,fulga2012scattering,hu2015measurement}, e.g. with external leads. The second way was to probe the edge through the scattering of bulk waves, namely \emph{inside} the sample, at the boundary \cite{graf2013bulk,bal2017topological}. Our strategy is to apply the latter approach to continuous models in order to explore the asymptotic part of the spectrum, {hence revealing the possible presence of ghost edge modes}.

Note that a different way to study the edge problem for continuous models is to consider a confining potential or a continuous interface between two topologically distinct samples \cite{fefferman2016edge,bal2017topological,bal2018continuous,faure2019manifestation,drouot2019bulk}. Such an interface is smoother than a sharp boundary and usually regularizes the problem so that there is no hidden mode at infinity. However, with a few exceptional cases, the counterpart of this approach is the loss of exact solvability. The main conclusion of this paper is that the bulk-edge correspondence for a sharp boundary is also perfectly valid as long as all edge modes, including the ones hidden at infinity, are properly taken into account.

The paper is organized as follows. In Section \ref{sec:themodel} we present a continuous model and compute the edge spectrum for different boundary conditions, revealing an apparent anomaly. Section \ref{sec:anomalousBEC} discusses the bulk-edge correspondence in details in order to quantify the previous mismatch. Section \ref{sec:scattering} introduces scattering theory and solves the paradox. Section \ref{sec:Dirac} shows the universality of this approach by applying it to the Dirac Hamiltonian. Section \ref{sec:discussion} concludes and suggest several consequences of this new paradigm. 

\section{Shallow-water with odd viscosity \label{sec:themodel}}

The two-dimensional rotating shallow-water model, linearized around a rest state in a rotating reference frame, is ruled by the following system:
\begin{subequations}\label{eq:general_model}
\begin{align}
\label{eq:general_model_eta}\partial_t \eta &= - \partial_x u - \partial_y v\\
\label{eq:general_model_u}\partial_t u &=-\partial_x \eta+ \left(f +\epsilon \nabla^2 \right)v \\
\label{eq:general_model_v}\partial_t v &=-\partial_y \eta- \left(f +\epsilon \nabla^2 \right)u 
\end{align}
\end{subequations}
where $(u,v)$ are the two velocity components in the plane $(x,y)$, $\eta$ the interface elevation relative to the mean depth $H=1$, $f$ the Coriolis parameter and $\epsilon$ the odd viscosity parameter. Time unit has been chosen such that phase speed is $\sqrt{gH}=1$, with $g$ the standard gravity. In the absence of off viscous terms (when $\epsilon=0$ above) it was realized that equatorial waves on Earth could be interpreted as topological modes of this flow when $f$ varies with $y$ and changes sign at the equator \cite{delplace2017topological}. In what follows we consider both $f$ and $\epsilon$ positive and homogeneous in space.
For geophysical fluids $\epsilon$ is nothing but an arbitrarily small regularisation parameter, in contrast to active matter systems described by a similar model and where $\epsilon$ can be tuned to large values. Indeed this models occurs in various context beyond geophysical fluids \cite{souslov2018topological,van2018photon} and appears as a paradigmatic two-dimensional model with 3 bands and spin-1 symmetry, by analogy with the Dirac Hamiltonian that has 2 bands and spin-1/2 symmetry. We also discuss the latter in detail in Section \ref{sec:Dirac}.

\subsection{The bulk picture \label{sec:bulk_picture}}

We briefly recall some known facts about the the bulk problem, where $(x,y) \in \mathbb R^2$. We look for normal modes of the form $(\eta,u,v) = (\hat\eta,\hat u, \hat v) \ee^{\ii (\omega t - k_x x - k_y y)}$ leading to the eigenvalue problem
\begin{equation}\label{eq:bulk_hamiltonian}
 \omega \begin{pmatrix}
 \hat \eta \\ \hat u \\ \hat v 
 \end{pmatrix}=
 \begin{pmatrix}
 0 & k_x & k_y \\ k_x & 0 & -\ii (f-\epsilon  k^2) \\ k_y & \ii (f-\epsilon  k^2) & 0
 \end{pmatrix} \begin{pmatrix}
 \hat \eta \\ \hat u \\ \hat v 
 \end{pmatrix}
\end{equation}
There are three bands: $\omega_\pm = \pm\sqrt{k^2 + (f-\epsilon k^2)^2}$ with $k^2 = k_x^2+k_y^2$ and $\omega_0 = 0$. These band will be reminiscent in the edge picture, see below. In particular the system is gaped for $f \neq 0$ and each band has a well-defined topological invariant: the Chern number. Respectively $C_\pm=\pm 2$ and $C_0=0$ for $f>0$ and $\epsilon>0$. Each non-vanishing Chern number captures a twist in the corresponding eigenfunction $(\hat\eta_\pm,\hat u_\pm, \hat v_\pm)$ as $(k_x,k_y)$ varies over $\mathbb R^2$. It is actually not well-defined for $\epsilon = 0$ and it was realized recently that odd-viscosity ensures that the bulk problem is properly regularised \cite{souslov2018topological,tauber2018odd}. This is analogous to the regularization of Dirac Hamiltonian \cite{volovik1988analogue,bal2018continuous} (see also Section \ref{sec:Dirac}). The main issue that remains is the regularization of the edge picture.

\subsection{The edge picture \label{sec:edge_picture}}

\begin{figure}[htbp]
\centering
\includegraphics[scale=0.65]{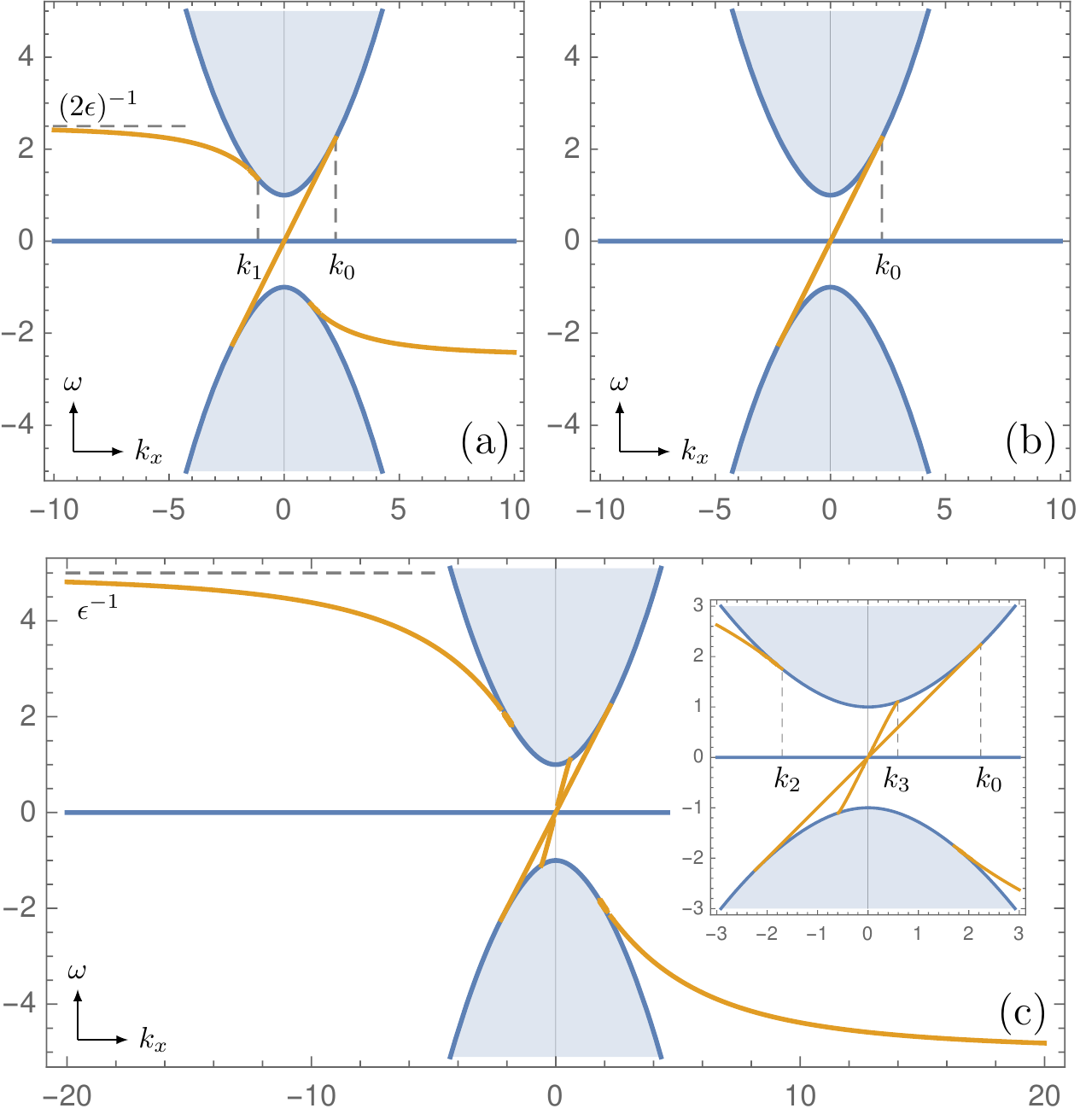}
\caption{Edge modes confined at the boundary $y=0$ for $f=1$, $\epsilon=0.2$ and three different boundary conditions, according to \eqref{eq:boundary_conditions}: (a) DD (b) DM (c) DS. In all cases the Kelvin wave is present, with linear dispersion relation and $|k_x|<k_0$. For DM this is the only mode. For DD (resp. DS) one has an extra mode merging into the upper band at $k_1$ (resp $k_2)$ and saturating at $\omega = 1/(2\epsilon)$ (resp. $\epsilon$). For DS there is a third mode with almost linear dispersion relation $\omega \sim 2k_x$ and merging with the bulk at $k_3$. The blue curves delimits the region of the (projected) bulk bands. \label{fig:edge_modes} }
\end{figure}

In the edge picture, where $(x,y) \in \mathbb R\times \mathbb R^+$, we study three boundary conditions that are relevant for the topological aspects:
\begin{subequations}\label{eq:boundary_conditions}
\begin{align}
\label{eq:boundary_condition_DD}\text{DD:} \qquad v(y=0) = 0, \quad \& \quad & u(y=0) = 0, \\
\label{eq:boundary_condition_DM}\text{DM:} \qquad v(y=0) = 0, \quad \& \quad  & (\partial_x u + \partial_y v)|_{y=0} = 0, \\
\label{eq:boundary_condition_DS}\text{DS:} \qquad v(y=0) = 0, \quad \& \quad & (\partial_x u - \partial_y v)|_{y=0} = 0. 
\end{align}
\end{subequations}
In the following we call \eqref{eq:boundary_condition_DD} Dirichlet-Dirichlet (DD), also called no-slip; \eqref{eq:boundary_condition_DM} is called Dirichlet-Membrane (DM) by noticing that from \eqref{eq:general_model_eta} it implies $\partial_t \eta =0$ at the boundary;  \eqref{eq:boundary_condition_DS} is called Dirichlet-Stressfree (DS) since it imposes a vanishing force by the boundary on the fluid. We stress that each boundary condition consist of two constraints only. In particular $\eta$ is not always constrained. Moreover not all the constraints are allowed because the self-adjointness of the problem has to be preserved. For example $u=0$ and $\eta=0$ at $y=0$ is not an adequate boundary condition. See Appendix \ref{app:self-adjoint} for a general rule of the allowed boundary conditions.

The system is invariant under translation in the $x$-direction so we look for normal modes of the form $(\eta,u,v) = (\hat\eta,\hat u, \hat v) \ee^{\ii (\omega t - k_x x)}$. Inserting it into \eqref{eq:general_model_eta} we realize that $\hat \eta = \omega^{-1}(k_x \hat u + \ii \partial_y \hat v)$ can be eliminated when inserted into \eqref{eq:general_model_u} and \eqref{eq:general_model_v}. We end up with a system of two ordinary differential equations of order two in $y$ and with constant coefficients, depending on the parameters $\omega$ and $k_x$:
\begin{align}
\label{eq:edge_ode_v}\big(\epsilon \partial_{yy}  - \dfrac{k_x}{\omega} \partial_y  + ( f - \epsilon k_x^2)\big) \hat v &= \dfrac{\ii}{\omega}(\omega^2-k_x^2) \hat u \\
\label{eq:edge_ode_u}\big(\epsilon \partial_{yy}  + \dfrac{k_x}{\omega} \partial_y  + ( f - \epsilon k_x^2)\big) \hat u &= -\dfrac{\ii}{\omega}(\partial_{yy}+\omega^2) \hat v
\end{align}
This problem is solvable analytically. We look for solutions that are confined near the boundary, namely such that $(u,v)\rightarrow 0$ as $y \rightarrow \infty$. In contrast to bulk normal modes, such solutions appear in the gaped region of the $(k_x,\omega)$-plane, complementary to the (projected) bulk bands. We first solve the general problem for any value of $k_x$ and $\omega$ in that region, then apply successively the different boundary conditions (DD, DM and DS). The details are provided in Appendix \ref{app:edge_problem} and the result is shown in Figure \ref{fig:edge_modes}. 

We observe that the number of modes in each gap, that is supposed to be topological, depends on the choice of the boundary condition. In each gap we respectively count 2, 1 and 3 modes for DD, DM and DS. Moreover we observe the presence of edge modes leaving a bulk band and saturating at some constant frequency $\omega \propto \epsilon^{-1}$, showing that the edge problem is not compactified at $k_x \rightarrow \infty$, even if the bulk is. Moreover, the way to count these edge modes correctly is also puzzling, but the total number can anyway not coincide with the Chern number as it depends on the boundary condition. The bulk-edge correspondence seems anomalous.

\section{Anomalous bulk-edge correspondence \label{sec:anomalousBEC}} 

We define in this section a precise number of edge modes to quantify properly the bulk-edge correspondence anomaly reported in the previous section. This is an essential step to solve the paradox in the next section. 

\subsection{Bulk-edge correspondence in condensed matter  \label{sec:standardBEC}}

\begin{figure}[htbp]
\centering
\includegraphics[scale=0.8]{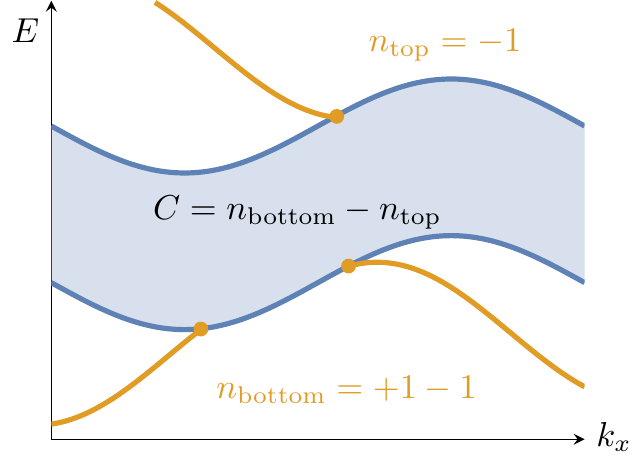}
\caption{The standard bulk-edge correspondence in its most general form.\label{fig:standard_BEC} The number of edge modes (orange) below and above the bulk band (shaded blue) is defined by the crossing with its external lines (solid blue). The sign depends if the mode is disappearing or emerging in the bulk band, with a relative global sign for the top and the bottom.}
\end{figure}

Consider a conventional band of a Hamiltonian ruling a two-dimensional system, e.g. a tight-binding model, with Chern number $C$. In the edge picture (half-plane geometry with boundary at $y=0$), the projection of this band may be connected to edge modes coming from the gap above and below, as illustrated in Figure \ref{fig:standard_BEC}. As $k_x$ is increasing, these modes can disappear into the band or emerge from it. For the bottom of the band, we \emph{define} the number of edge modes $n_\mathrm{bottom}$ as the algebraic counting of the points where an edge state disappears ($+1$) or emerges ($-1$). For the top of the band we define similarly $n_\mathrm{top}$, except that the signs are inverted \footnote{Edge numbers $n_\mathrm{top}$ and $n_\mathrm{bottom}$ are defined up to a global sign, depending on the orientation of the boundary, but the relative sign between them in their definition persists anyway.}. 
This is equivalent to count the number of crossing of edge modes with the external lines of the bulk band, with a sign depending on the dispersion relation $\tfrac{\partial E}{\partial k_x}$ at the crossing. The bulk-edge correspondence is given by \cite{hatsugai1993chern}
\begin{equation}\label{BEC}
C = n_\mathrm{bottom}-n_\mathrm{top}
\end{equation}
Moreover if the problem satisfies a further assumption, quite common in condensed matter, this correspondence can be rewritten in a simpler form. Consider a system with $N$ bands denoted by $i \in \{1,\ldots, N\}$, ordered by increasing energy and separated by spectral gaps, the corresponding topological numbers are $C_i$, $n^i_\mathrm{top}$ and $n^i_\mathrm{bottom}$. If we assume that both $k_x$ and $H$ are bounded (e.g. in a tight-binding model with a Brillouin zone)
then necessarily $n^i_\mathrm{top} = n^{i+1}_\mathrm{bottom} :=n^i$ for $1 \leq i \leq N-1$ so that there is only one edge invariant $n^i$ per gap $i$ above the band $i$, that can be computed by the algebraic crossing with a horizontal line inside the gap (e.g. constant Fermi level). Moreover $n^1_\mathrm{bottom} = 0$ and $n^N_\mathrm{top} = 0$. In that case the correspondence can be rewritten $n^i = -\sum_{j=1}^i C_i$, namely the number of edge modes in a gap is given by the sum of the Chern numbers of all band below it (up to a global sign depending of the orientation of the boundary) \cite{hatsugai1993chern}. However we claim that this relation is less general than \eqref{BEC}, the latter being still satisfied when the previous assumption is not.

\subsection{Anomaly in the continuous model}
In the continuous model from Section \ref{sec:themodel} neither $k_x$ nor $\omega$ (analogue to $E$) are bounded so that the aformentioned assumption is not satisfied. We can however define a precise number of edge modes for each boundary condition, even for the modes that saturates asymptotically at a constant $\omega$. This is summarized in Table \ref{tab:edge_modes}.

\begin{table}[htbp]
    \centering
    \begin{tabular}{|c|c|c|c|}
    \hline
       Boundary condition  & DD & DM & DS \\
       \hline\hline
        $n^+_\mathrm{bottom}$ & 2 & 1 & 3\\\hline
        $n^0_\mathrm{top}$    & 1 & 1 & 2\\\hline
        $n^0_\mathrm{bottom}$ & 1 & 1 & 2\\\hline
        $n^-_\mathrm{top}$    & 2 & 1 & 3 \\\hline
    \end{tabular}
    \caption{The number of edge modes around each band for different boundary conditions.}
    \label{tab:edge_modes}
\end{table}

The middle band is never anomalous since $n^0_\mathrm{top} = n^0_\mathrm{bottom}$ regardless of the boundary condition, which is compatible with \eqref{BEC} and $C_0=0$. Moreover we notice that $n^0_\mathrm{top}\neq n^+_\mathrm{bottom}$ although it corresponds to the same gap between the middle and the upper band, but this is not a problem for the bulk-edge correspondence \eqref{BEC}, since it focuses on a specific band rather than a gap. However the upper and lower band are anomalous: they are not bounded so the numbers $n^+_\mathrm{top}$ and $n^-_\mathrm{bottom}$ make no sense. If we naively set them to $0$, then the bulk-edge correspondence is satisfied for DD boundary condition: $C_+ = n^+_\mathrm{bottom} =2$ and $C_- = -n^-_\mathrm{top} =-2$, but we see immediately that the boundary conditions DM and DS are anomalous. 

Nevertheless we claim that the bulk-edge correspondence \eqref{BEC} still makes sense, and the purpose of the next section is to provide a more general definition of the edge numbers, allowing for an explicit computation of $n^+_\mathrm{top}$ and $n^-_\mathrm{bottom}$ and so that \eqref{BEC} is restored for each band and any boundary condition.

\section{Scattering theory \label{sec:scattering}}

In this section we provide an alternative formalism to define and compute the number of edge modes above and below each band. As we shall see it reproduces the result from Table \ref{tab:edge_modes} independently, but it also allows for a definition of (generalized) edge modes at infinite $\omega$, so that the bulk-edge correspondence \eqref{BEC} is recovered.

\begin{figure}[htb]
	\centering
\includegraphics[scale=0.65]{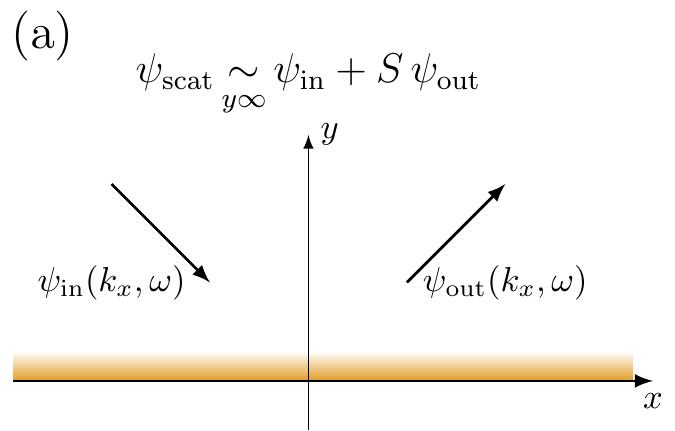}
\includegraphics[scale=0.6]{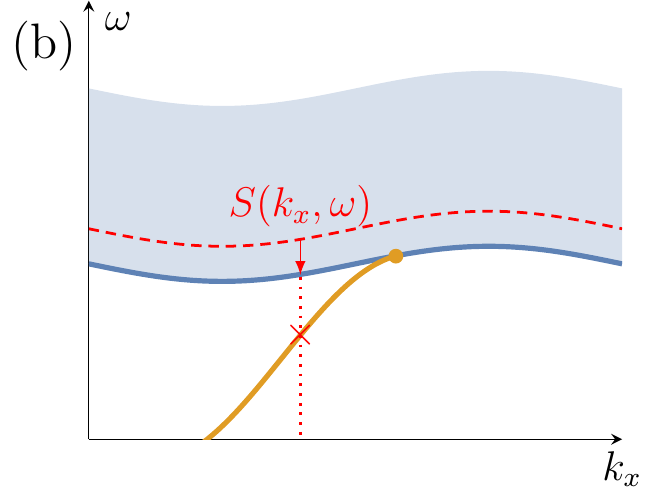}
\caption{The concept of scattering theory. (a) A linear combination, encoded by $S$, of incoming and outgoing bulk states is solution to the edge problem. (b) The scattering matrix (here a $U(1)$-phase) is defined for $k_x$ and $\omega$ in the projected bulk band. When approaching its extremity, the argument of $S$ counts the number of bound states below: the edge modes in our case. As $k_x$ varies, the relative argument of $S$ counts the number of edge modes that have vanished (or emerged) in the band.\label{fig:intro_scattering}}
\end{figure}

The formalism of scattering theory was developed in \cite{graf2013bulk} to prove the bulk-edge correspondence for tight-binding models of condensed matter. We first review the general concepts involved and implement them explicitly in our case. The scattering matrix $S$ encodes how bulk waves, that propagate \emph{inside} the sample, are reflected at its edge (Figure \ref{fig:intro_scattering}(a)). The normal modes from Section \ref{sec:bulk_picture} are not solution to the boundary problem from Section \ref{sec:edge_picture}, but a linear combination of an incoming state $\psi_\mathrm{in}$ and an outgoing state $\psi_\mathrm{out}$ can be. The scattering matrix $S$ is then defined as the relative coefficient between these two states. See the precise definition below.

The interest of $S$ resides in the application of Levinson's theorem \cite{graf2013bulk}. At fixed $k_x$ and for $\omega \rightarrow \omega_\mathrm{min}(k_x)$, the bottom of the bulk band (when it exists), the argument of the scattering matrix is equal to the number of bound states below it (Figure \ref{fig:intro_scattering}(b)). In this context, they are precisely the edge modes that could appear below the bulk band. Then, as $k_x$ increases, the argument of $S$ stays the same until an edge mode disappears in (resp. emerges from) the bulk band, in which case the argument changes by $2\pi$ (resp $-2\pi$). The number of edge modes between $k_1$ and $k_2$ is thus counted by \cite{graf2013bulk}
\begin{equation}
n = \lim_{\varepsilon \rightarrow 0} \dfrac{1}{2\pi} \mathrm{Arg}[S(k_x,\omega_\mathrm{min}(k_x)+\varepsilon)]\big|_{k_1}^{k_2}
\end{equation}
Note that a similar discussion is valid for the upper limit of the band (when it exists), up to a global sign. For usual condensed matter systems, we take $k_2=k_1+2\pi$, namely a full loop over the reduced Brillouin zone, so that we get $n=n_\mathrm{bottom}$ from Section \ref{sec:standardBEC}. In our case we will take $k_1 \rightarrow -\infty$ and $k_2 \rightarrow \infty$. 

\subsection{The scattering matrix} To define $S$
we recall some data from the bulk. For the rest of the discussion we focus on the upper bulk band since the lower one can be studied in an analogous way. The normal mode associated to \eqref{eq:bulk_hamiltonian} and $\omega = \omega_+$  is $(\eta,u,v) = \hat \psi \ee^{\ii (\omega t - k_x x - k_y y)}$ with
\begin{equation}\label{eq:defPsi+}
\hat \psi(k_x,k_y)  = \dfrac{1}{\sqrt 2\, k} \begin{pmatrix}
k^2\, /\omega_+(k) \\ k_x - \ii k_y (f-\epsilon k^2) / \omega_+(k)\\k_y + \ii k_x  (f-\epsilon k^2)/ \omega_+(k) 
\end{pmatrix}
\end{equation}
This family is singular at $k=0$ and $k\rightarrow \infty$ but each singularity can be removed up to a gauge transformation: $\hat \psi_{0/\infty} := \lambda_{0/\infty} \psi$ where $\lambda_0 = k^{-1}(k_x + \ii s_f k_y)$ and $\lambda_\infty = k^{-1}(k_x - \ii s_\epsilon k_y)$ are $U(1)$-phases ($s_f$ and $s_\epsilon$ are the respective sign of $f$ and $\epsilon$), see \cite{tauber2018odd}. In the following we shall consider $\hat \psi_0$ or $\hat \psi_\infty$ according to the region we are looking at. 

In the edge picture, we fix $\omega > f$ and $k_x$ in the projected bulk band and away from the singular points, and denote $k_y := \kappa$ to emphasize that it is not conserved. In the bulk, it is a fact that the equation $\omega_+(k_x,\kappa) = \omega$ always has at least two real solutions in $\kappa$, and possibly other solutions with non-vanishing imaginary part \cite{graf2013bulk}. In our case, $\omega^2 = k_x^2  + \kappa^2 + (f - \epsilon(k_x^2  + \kappa^2))^2$ has four solutions in $\kappa$, that we denote by $\kappa_\mathrm{in/out} = \mp \sqrt{K_+}$ and $\tilde \kappa_{\pm} = \pm \ii \sqrt{-K_-}$ where
\begin{equation}\label{eq:Kpm}
K_\pm = \dfrac{1}{2\epsilon^2} \Big(-(1-2\epsilon (f-\epsilon k_x^2)) \pm \sqrt{1 - 4 \epsilon f + 4 \epsilon^2 \omega^2} \Big)
\end{equation}
Indeed for $\omega$ and $k_x$ in the region of the upper bulk band, $K_+ \geq 0$ and $K_-\leq 0$ so that $\kappa_\mathrm{in/out} \in  \mathbb R$ and $\tilde \kappa_{\pm} \in \ii \mathbb R$. Since $\tfrac{\partial \omega_+}{\partial \kappa}<0$ for $\kappa = \kappa_\mathrm{in}$ then $\psi_\mathrm{in} := \hat \psi(k_x,\kappa_\mathrm{in}) \ee^{\ii(\omega_+t-k_xx -\kappa_\mathrm{in} y)}$ is an incoming normal mode at frequency $\omega_+$. Similarly $\kappa_\mathrm{out}$ describes an outgoing mode $\psi_\mathrm{out}$. The two other solutions describe modes that are exponentially increasing and decreasing away from the boundary $y=0$. One of them is allowed and is a bound state, namely $\psi_\mathrm{b} :=  \hat \psi(k_x,\tilde \kappa_-) \ee^{\ii(\omega_+t-k_xx)}\ee^{-|\tilde \kappa_-| y}$. This state is actually necessary to satisfy non-trivially the constraints of a boundary condition. The scattering state is defined by
\begin{equation}
\psi_\mathrm{scat}(x,y,t) =  \alpha \psi_\mathrm{in} + \beta \psi_\mathrm{out} + \gamma \psi_\mathrm{b}
\label{eq:scattering}
\end{equation}
with $\alpha,\beta$ and $\gamma$ are coefficients that depends on $k_x$ and $\omega$ which are adjusted to satisfy the boundary condition at $y=0$, so that $\psi_\mathrm{scat}$ is a solution of the edge problem as a superposition of bulk solutions. The scattering matrix is 
\begin{equation}
S(k_x,\omega) := \dfrac{\beta}{\alpha}
\label{eq:scattering_matrix}
\end{equation}
In our case the eigenspace is of dimension 1, so that $S \in U(1)$ (the unitarity is ensured by a proper normalization of the scattering state \cite{graf2013bulk}).

\subsection{Bottom band scattering} 
\begin{figure}[htb]
	\centering
\includegraphics[scale=0.7]{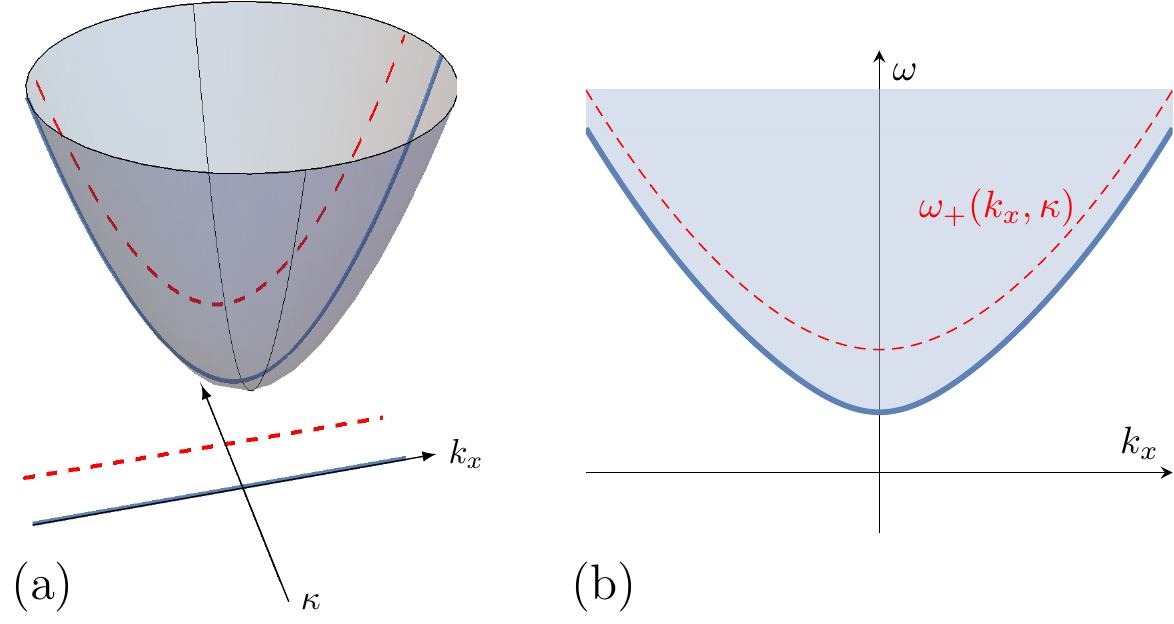}
	\caption{(a) 3D Plot of the bulk band $\omega_+(k_x,\kappa)$. The red dashed curve is $\omega_+(k_x,\kappa)$ for fixed $\kappa$ and the blue one is for $\kappa=0$. (b) Projection of $\omega_+$ in the edge picture. We shall look at the winding number of the scattering matrix as $k_x$ varies along the dashed red curve, namely for fixed $\kappa$, and then take the limit $\kappa \rightarrow 0$. \label{fig:scattering_kappa}}
\end{figure}

We would like to look at the scattering matrix along the bottom of the band $\omega_+$ instead of a fixed $\omega$. In the edge picture the bulk band is projected: for fixed $\kappa$, $\omega_+(k_x,\kappa)$ describes a curve into the bulk band region that goes to the bottom of it when $\kappa \rightarrow 0$, see Figure \ref{fig:scattering_kappa}. 

Thus we consider the scattering problem at fixed $k_x$ and $\kappa$, the latter being small, and $\omega = \omega_+(k_x,\kappa)$. Then we look at the winding number as $k_x$ varies, and eventually take the limit $\kappa \rightarrow 0$. We now set $\kappa_\mathrm{in} = -\kappa <0$, and deduce from \eqref{eq:Kpm} and definitions of $\kappa_\mathrm{out}$ and $\tilde \kappa_\pm$ below it that $\kappa_\mathrm{out} = - \kappa_\mathrm{in} =  \kappa$ and
\begin{equation}
 \tilde \kappa_- = -\ii \sqrt{\kappa^2 + \dfrac{1}{\epsilon^2}(1-2\epsilon(f-\epsilon k_x^2))}
\end{equation}
In particular, notice that $\lim \tilde \kappa_- \neq 0$ as $\kappa \rightarrow 0$. The scattering state becomes
 \begin{align}\label{eq:psi_scat_bottom}
 \psi_\mathrm{scat}(y) = & \alpha \hat \psi_0(k_x,-\kappa)\ee^{\ii \kappa y} + \beta \hat \psi_0(k_x,\kappa) \ee^{-\ii \kappa y} \cr & + \gamma \hat \psi_0(k_x,\tilde \kappa_-(k_x,\kappa)) \ee^{- |\tilde \kappa_-(k_x,\kappa)| y}
 \end{align}
We dropped the $x$ and $t$ dependence that is trivial, and used $\hat \psi_0$ that is regular around $k_x,\kappa=0$. Then we impose a boundary condition from \eqref{eq:boundary_conditions}, that will constraint two of the three parameters $\alpha, \beta$ and $\gamma$, allowing a non-ambiguous definition of $S(k_x,\kappa) = \beta/\alpha$. Note that this is not a coincidence: the number of conditions required at the boundary is deeply related to the number of solutions $\kappa$ to $\omega_+(k_x,\kappa)=\omega$, which fixes the number of free parameters in the scattering states \cite{graf2013bulk}.

\begin{figure}[htbp]
\centering
\includegraphics[scale=0.65]{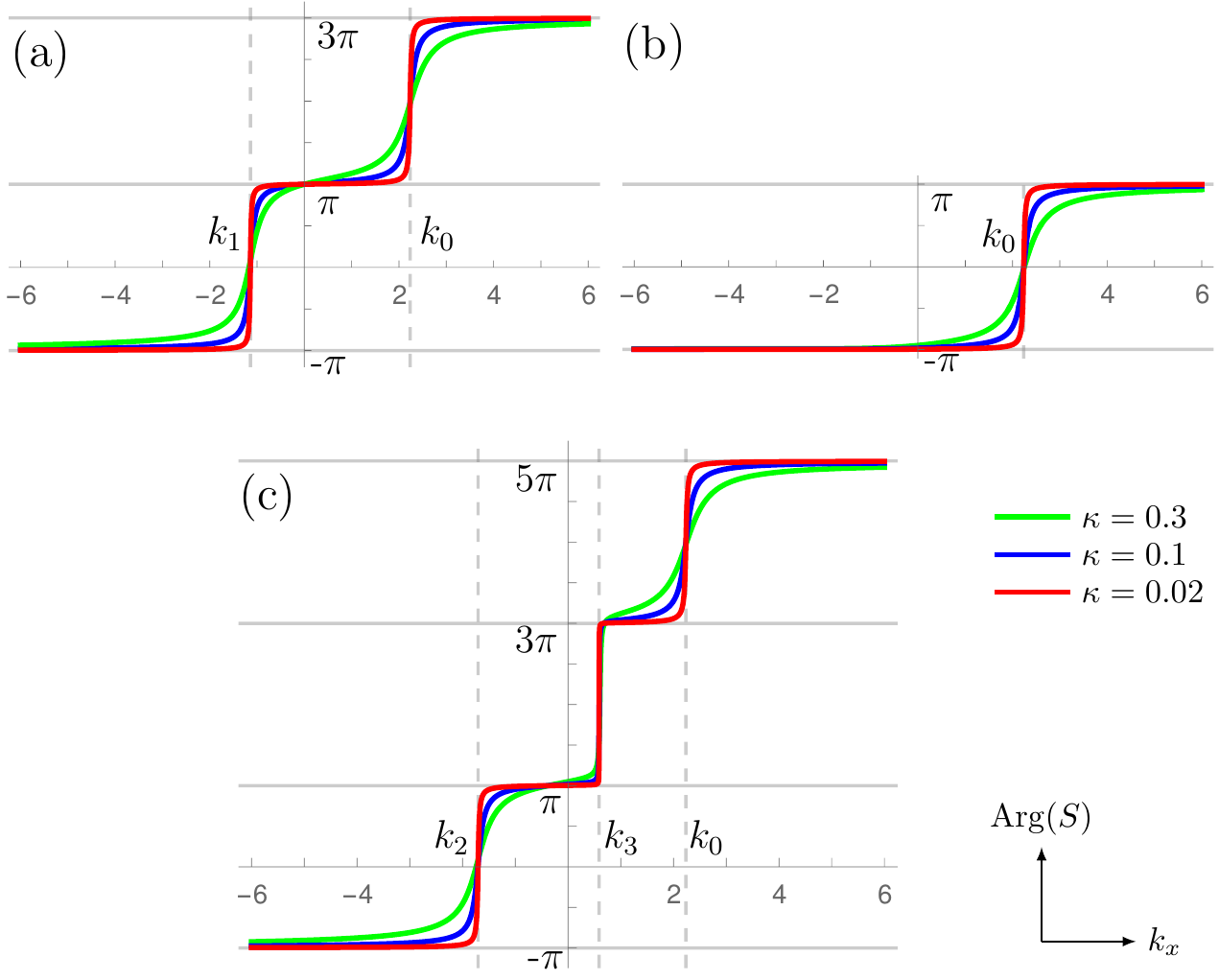}
\caption{Argument of $S$ at the bottom of the band $\omega_+$ for $f=1$, $\epsilon =0.2$ and $\kappa=0.3,\,0.1$ and $0.02$ (respectively in green, blue and red) for different boundary conditions. For DD (a), DM (n) and DS (c), the winding number of $S$ is -2,-1 and -3, in agreement with $n^+_\mathrm{bottom}$ from Table \ref{tab:edge_modes}. The points $k_x=k_i$, where the jumps occur in the $\kappa \rightarrow 0$ limit, are the same than in Figure \ref{fig:edge_modes} where the edge modes merge into the bulk band. \label{fig:scattering_bottom} }
\end{figure}

For each boundary condition in \eqref{eq:boundary_conditions} we can define and compute $S \in U(1)$ and look at its complex argument at the bottom of $\omega_+$, namely when $k_x$ varies from $-\infty$ to $+ \infty$ and $\kappa \rightarrow 0$. The scattering data is detailed in Appendix \ref{app:scat_data} and the argument of $S$ is plotted in Figure \ref{fig:scattering_bottom}. We observe that the winding number of $S$ is $w^+_\mathrm{bottom} = 2,\,1$ and $3$, respectively for DD, DM and DS, in agreement with $n^+_\mathrm{bottom}$ from Table \ref{tab:edge_modes}. Moreover, as $\kappa \rightarrow 0$, the jump of $\mathrm{Arg}(S)$ occurs precisely at the points $k_x=k_i$ ($i=0,\ldots,3)$ where the edge modes merge into the bulk band, compare with Figure \ref{fig:edge_modes}. 

\subsection{Infinite top band scattering} 
\begin{figure}[htbp]
\centering
\includegraphics[scale=0.7]{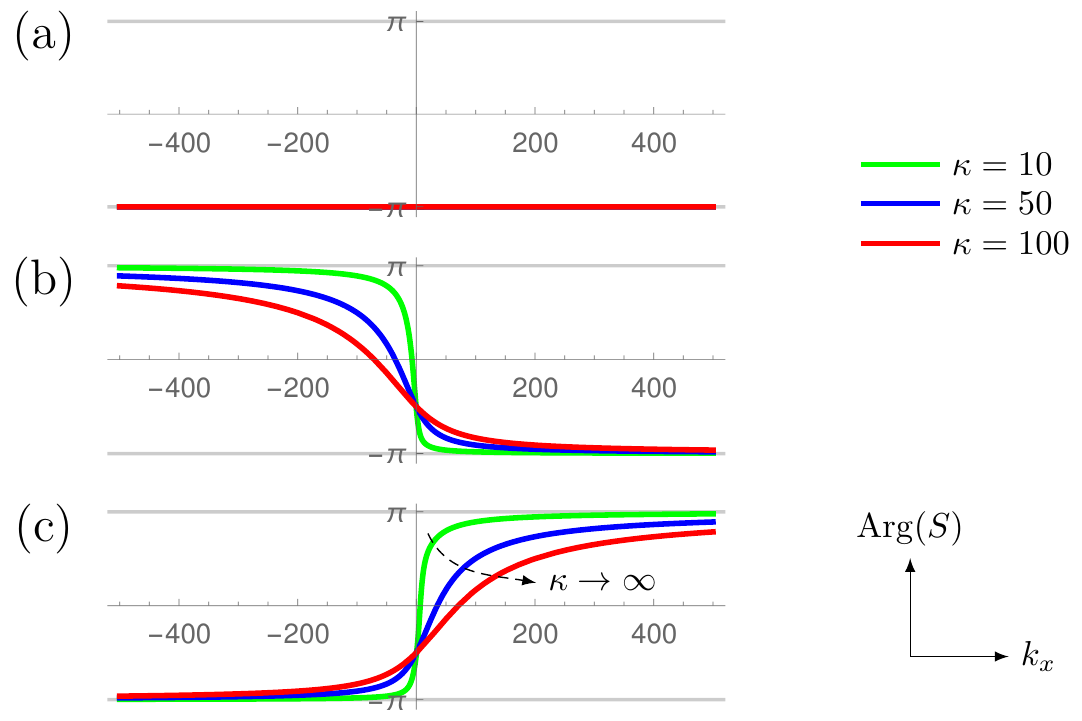}
\caption{Argument of $S$ at the (infinite) top of the band $\omega_+$ for $f=1$, $\epsilon =0.2$ and $\kappa=10,\,50$ and $100$ (respectively in green, blue and red) with different boundary conditions. For DD (a) the three curved are superposed to the constant value $\pi$, so that $S$ does not wind. For DM (b) and DS (c) one has a non-vanishing winding number: respectively -1 and 1. Note that the red curve is the closest to the $\kappa \rightarrow \infty$ limit, so that this winding is delocalized in $k_x$, rather than converging to a localized jump, in contrast to the scattering at the bottom of the band (Figure \ref{fig:scattering_bottom}).\label{fig:scattering_top} }
\end{figure}

As we have seen the scattering formalism provides an alternative way to compute the number of (standard) edge modes below the band, that is consistent with the method from Section \ref{sec:anomalousBEC}. However, it is more general than the latter because it allows to count the number of edge modes at the top of the band, even if the upper band is not bounded from above. Indeed we simply compute the scattering matrix as before, but instead we take $\kappa \rightarrow \infty$, which corresponds to the (infinite) edge of the upper band. Moreover in that case we are near the $k \rightarrow \infty$ point that may be singular, so we compute the scattering data with $\hat \psi_\infty$ that has no singularity there, instead of $\hat \psi_0$. This is done in Appendix \ref{app:scat_data} and the argument of $S$ is plotted in Figure \ref{fig:scattering_top}. 

We observe that $S$ has a well-defined winding number for DD, DM and DS as we explore the upper limit of the band: respectively 0, -1 and 1. Moreover we stress that in this limit the argument of $S$ is \emph{not} converging to a localized jump  but rather completely delocalized in $k_x$, so that one has to explore the whole parameter $k_x \in \mathbb R$ in order to compute it. Finally we call this winding number $n^+_\infty$, which we interpret as the number of edge modes at the (infinite) top of the band $\omega_+$. If we compare it with the edge number at the bottom of the band, we conclude that the bulk-edge correspondence \eqref{BEC} is not anomalous anymore, namely the difference between the two numbers always gives the Chern number of the upper band (see Table \ref{tab:upper_band}). 

\begin{table}[htb]
    \centering
    \begin{tabular}{|c|c|c|c|}
    \hline
       Boundary condition  & DD & DM & DS \\
       \hline\hline
       $n^+_\infty$ & 0 & -1 & 1 \\\hline
        $n^+_\mathrm{bottom}$ & 2 & 1 & 3\\\hline
        $C_+$ & 2 & 2& 2 \\\hline
    \end{tabular}
    \caption{The number of (generalized) edge modes for the upper band. The bulk-edge correspondence \eqref{BEC} is properly satisfied regardless of the boundary condition if we identify $n^+_\infty=n^+_\mathrm{top}$.}
    \label{tab:upper_band}
\end{table}

\subsection{Inertial-like edge modes at infinity}

The scattering matrix detects the presence of edge modes at infinite frequency {that we dub \textit{ghost modes}}. \textit{A posteriori}, we can actually see a footprint of these modes  by exploring perturbatively the asymptotic regions of the gap in the limit of large wave number $|k_x|$. Let us consider $\epsilon>0, f>0$ and let us assume $\omega = \alpha |k_x|^\beta$ for some $1<\beta<2$ and $\alpha>0$ (i.e. below the band $\omega_+$ when   $k_x \rightarrow \pm \infty$). At the leading order in $k_x$,  the solutions localized near the edge are of the form (see \eqref{eq:def_Spm} and \eqref{eq:def_lambda})
\begin{align}
    & v(y) \sim V_3 \ee^{s_+ y} + V_4 \ee^{ s_- y},  \\
    & u(y) \sim -\ii \big(V_3 \ee^{s_+} - V_4 \ee^{s_- y}\big)
    \end{align}
    where
    $s_{\pm}=-|k_x|\left(1\pm\frac{\alpha}{2\epsilon } |k_x|^{\beta-2}\right)$.
These solutions are superpositions of \textit{inertial}-like waves, defined as waves with polarization relation $(\eta,u,v)=(0,1,\pm \ii)$. In the absence of odd viscosity, these waves are constant frequency modes $\omega=\pm f$, hence their name inertial. Because the odd viscous terms added into the problem have the structure of the Coriolis force (but depending on the wavenumber), it is not surprising that we recover such states at large  wavenumbers. 

The possible existence of a solution is discussed by applying the different boundary conditions. In the three cases considered above, the impermeability constraint $v(0)=0$ leads to $V_4=-V_3$.
  Thus for DD \eqref{eq:boundary_condition_DD} the second condition $u(0)=0$ leads to $V_3=0$ so that there is no asymptotic mode, in agreement with $n_\infty^+=0$. However for DM \eqref{eq:boundary_condition_DM} we get from \eqref{eq:membrane_lambda_s} the condition $ 2 k_x  = -(\alpha/\epsilon)|k_x|^{\beta-1}$ to have $V_3 \ne 0$. For $k_x \rightarrow -\infty$, there is a solution when $\beta \rightarrow 2$ and $\alpha=2\epsilon$. Instead, for $k_x \rightarrow +\infty$, there is no solution. 
 This indicates the presence of an  edge mode in the asymptotic upper-left region of the spectrum, whereas upper-right is empty. That is consistent with Figure \ref{fig:scattering_top}(b) where the jump of the argument seems to be ``pushed'' to $k_x \rightarrow -\infty$ as $\kappa \rightarrow +\infty$. Thus the scattering matrix counts the mismatch in the number of modes between $k_x = -\infty$ and $+\infty$. In this picture it seems that one mode has merged from the right to the ``top'' of the band, in agreement with $n_\infty^+ = -1$. Conversely, for DS the asymptotic expansion indicates the presence of a mode in the upper-right region, in agreement with Figure \ref{fig:scattering_top}(c) and $n_\infty^+ = +1$. 
 
 Interestingly, in the context geophysical fluid dynamics, an interpretation of the dispersion relation in shallow-water models with different boundary conditions  was proposed by Iga \cite{iga1995transition}, also by considering different asymptotic regimes in $(k_x,\omega)$ diagram.  In these regimes the initial problem is simplified and more tractable. Using an argument based on the conservation of the eigenfunction's zeros when $k_x$ is varied, Iga predicted the global shape of the spectra \cite{iga1995transition}, and generalized this method to other geophysical flow models \cite{iga2001transition}. This method gives robust information on the spectrum, such as the existence of modes that transit from one band to another when $k_x$ is varied (spectral flow), under fairly general assumptions (channel or cylinder geometry, parameters enforcing the existence of discrete spectrum,...). Here we have provided a complementary point of view using topology, where, again, asymptotic regions of the $(k_x,\omega)$ diagram must be taken into account to understand to the global shape of the spectrum.

\section{Dirac Hamiltonian \label{sec:Dirac}}

The choice of the shallow-water model was made here to illustrate the consequences on coastal waves in classical fluids, but our analysis of the bulk-edge correspondence applies to any two dimensional continuous model, as long as the bulk problem is properly compactified.

Postponing a general rigorous theorem to future work, we illustrate the power of our approach by applying the scattering formalism to the celebrated (massive) Dirac Hamiltonian, regularized by a $\epsilon k^2$ mass term
\begin{equation}
\label{eq:Dirac_regularized}
    H = \begin{pmatrix}
    m-\epsilon (\partial_x^2+\partial_y^2) & \ii \partial_x +\partial_y \\
    \ii \partial_x - \partial_y & -m+\epsilon (\partial_x^2+\partial_y^2)
    \end{pmatrix}.
\end{equation}
Such an Hamiltonian could describe for instance a two-dimensional $^3$He-$A$ superfluid phase, where the mass term $m$ would correspond to the chemical potential \cite{volovik1988analogue}.

When the mass $m$ is fixed, the presence of a regularization term $\epsilon k^2$ makes possible the introduction of well-defined Chern numbers of value
\begin{equation}
C_\pm = \pm\frac{\text{sign}(m)+\text{sign}(\epsilon)}{2} 
\end{equation}for the two eigenstates 
$\psi_\pm(k_x,k_y)$
of the bulk Hamiltonian 
\begin{equation}
    H_{\text{bulk}} = \begin{pmatrix}
    m -\epsilon k^2 & k_x - \ii k_y \\
    k_x + \ii k_y & -m +\epsilon k^2
    \end{pmatrix} \, .
    \label{eq:Dirac_bulk}
\end{equation} 
{with $k^2=k_x^2+k_y^2$,} that is derived from \eqref{eq:Dirac_regularized} by using a Fourier basis $\ee^{-\ii (k_x x + k_y y)}$ 
 (see Appendix \ref{sec:dirac} {and ref. \cite{bal2018continuous}}).

Let us then set $m$ and $\epsilon$ so that $C_+= 1$ and address the question of the boundary modes. 
For that purpose, we
consider two different boundary conditions {for $\psi:=(\phi_1,\phi_2)^T$} at $y=0$ that satisfy hermiticity (see Appendix \ref{app:self-adjoint})
\begin{subequations}\label{eq:boundary_conditions_Dirac}
\begin{align}
\label{eq:boundary_condition_Dirac1}\text{A:}& \quad \phi_1|_{y=0} = 0, \quad \& \quad  \phi_2|_{y=0} = 0, \\
\label{eq:boundary_condition_Dirac2}
\text{B:}&  \quad \phi_1|_{y=0} = 0, \quad \& \quad \partial_y \phi_2|_{y=0}  = -\ii \partial_x \phi_2|_{y=0}\,  
\end{align}
\end{subequations}

 The energy spectra for the boundary modes allowed by these two boundary conditions are derived  in Appendix  \ref{sec:dirac} and displayed in Figure \ref{fig:dirac} for $m=1$ and $\epsilon=0.1$.

\begin{figure}[htb]
\centering
\includegraphics[scale=0.65]{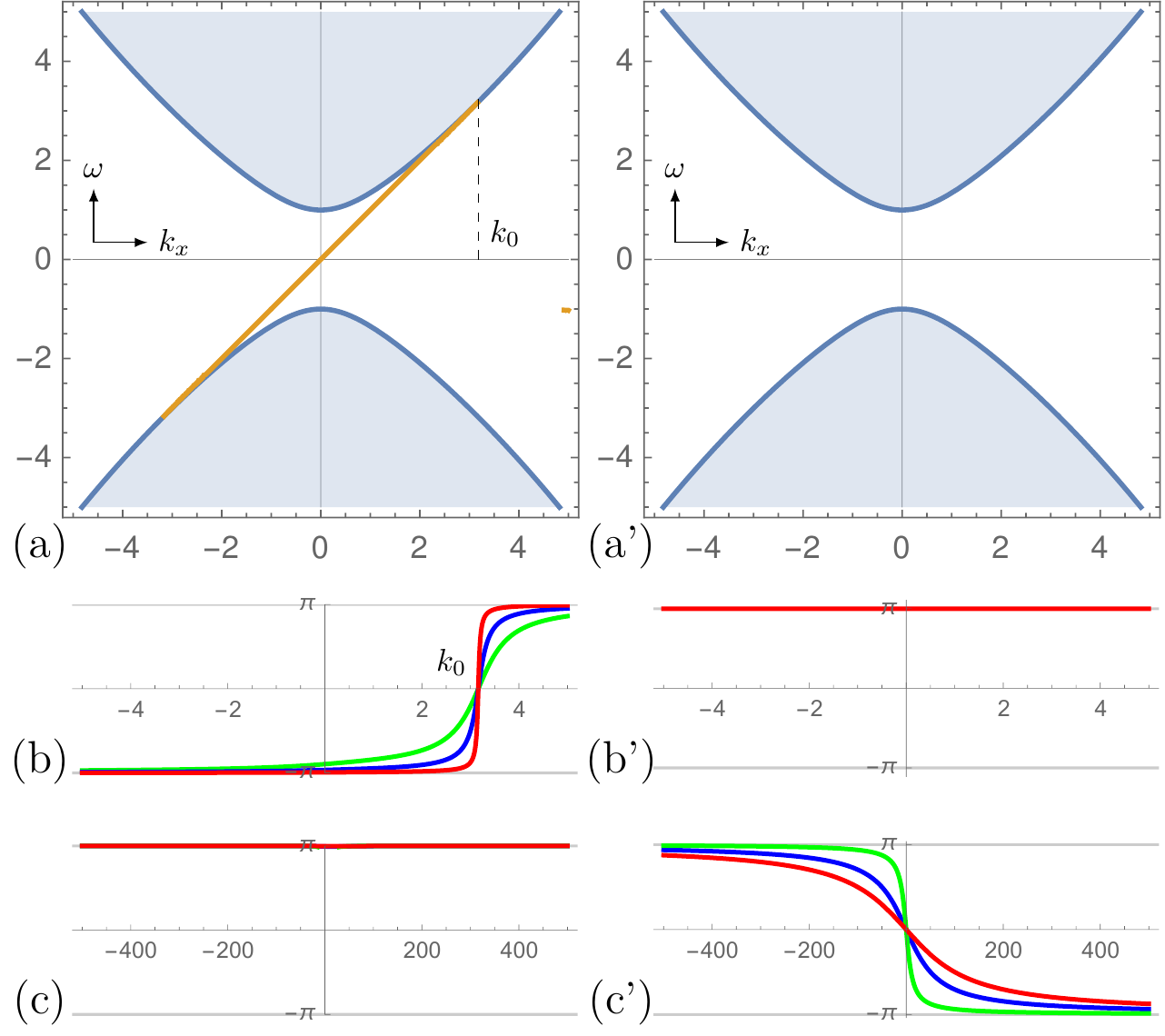}
\caption{{Dirac Hamiltonian with $m=1$ and $\epsilon=0.1$ (a) Edge spectrum with boundary condition A \eqref{eq:boundary_condition_Dirac1} (b) Arg(S) at the bottom of the upper band, for $\kappa = 0.3, 0.1$ and $0.02$ (respectively in green, blue red) (c) Arg(S) at the ``top'' of the upper band, for $\kappa = 10, 50$ and $100$ (respectively in green, blue red). (a'), (b') and (c') are the same plots for condition B  \eqref{eq:boundary_condition_Dirac2}} \label{fig:dirac} }
\end{figure}

Boundary conditions A yield the naively expected result from the values of the Chern numbers $C_{\pm} = \pm 1$, namely one chiral boundary mode that spans the bulk gap and propagates to the right (positive group velocity). The merging of this chiral mode into the bulk bands at $k_0 \approx \pm 3$ is well captured by the scattering theory introduced above and applied for the Dirac case in Appendix \ref{sec:dirac}. Figure \ref{fig:dirac} shows that this winding is indeed $+1$ for the top band, with a jump in phase that exactly occurs at $k=k_0$. It is also checked that no other evanescent state enters the band at $\omega \sim \infty$ ($n_\infty^+=0$), so that the winding number equals the Chern number and captures the number of modes gained by the bulk band.

In contrast, the boundary condition B does not allow boundary mode at finite energy and $k$. Accordingly, the winding number is zero  meaning that there is no evanescent mode entering the bulk bands. However, the winding $n_\infty^+=-1$ indicates the entrance of an \textit{ghost} boundary mode from the ``top'' of the band of positive energy, in agreement with the bulk-boundary correspondence, and the value of the Chern number.

\section{Discussion \label{sec:discussion}} 

To conclude, the apparent paradox of a mismatch in the bulk-edge correspondence for a continuous model with a sharp boundary is solved by the presence of {``ghost''} edge modes at infinity, that can be detected through the scattering formalism. Thus in continuous media the bulk-edge correspondence is always satisfied, independently from the boundary condition. This new paradigm can indeed be applied to any continuous model. Moreover it has various consequences and paves the way for new directions of investigation that we discuss now. \\

\begin{figure}[htb]
\centering
\includegraphics[scale=0.65]{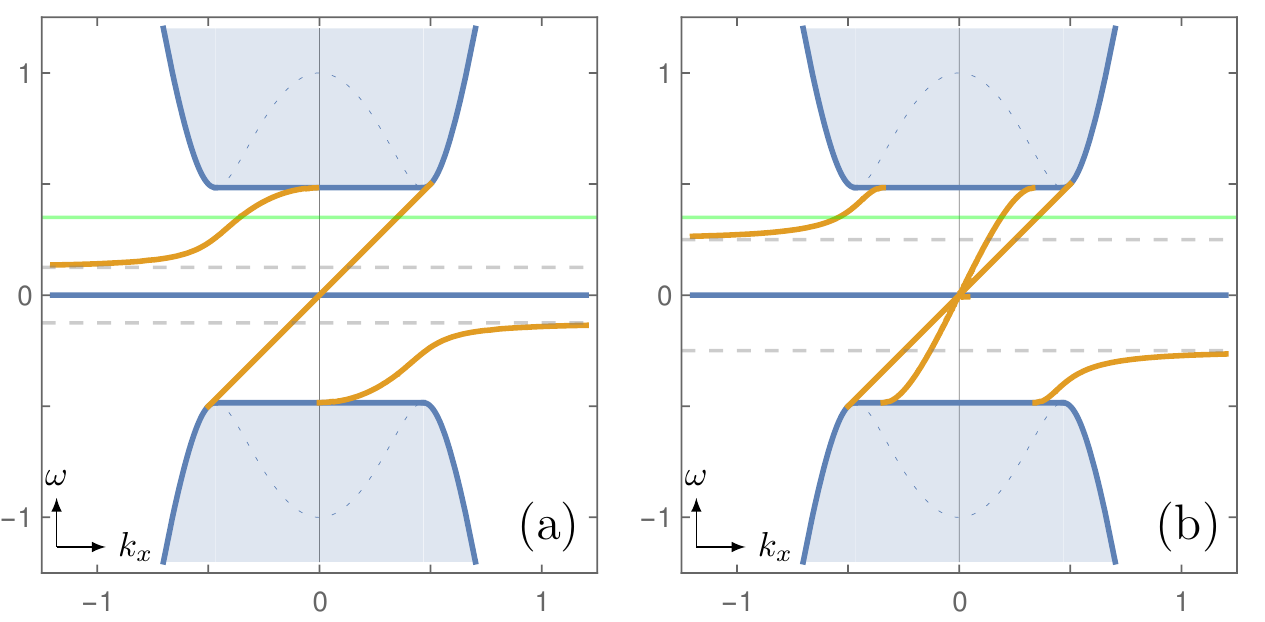}
\caption{Edge modes for $f=1$ and \emph{large} $\epsilon=4$, with two different boundary conditions according to \eqref{eq:boundary_conditions}: (a) DD and (b) DS. The number $n_\mathrm{cross}$, counting the crossing between the edge modes and the green horizontal line is respectively 2 and 3, but would be 1 and 2 in the case where $\epsilon$ is small (compare with Figure \ref{fig:edge_modes} (a) and (c)). Moreover, if we move the green line below some threshold $\propto \epsilon^{-1}$ (horizontal dashed line), or if we decrease $\epsilon$ continuously, then $n_\mathrm{cross}$ is also lowered by 1.  \label{fig:edge_modes_largeepsilon} }
\end{figure}

\paragraph{Contrary to a common belief, {chiral is not topological.}}
One usual way to define the edge number is to count the (algebraic) crossing $n_\mathrm{cross}(\omega)$ of the edge modes dispersion relation with a fiducial line $\omega = C^{te}$ in the gaped region (analogue to the Fermi energy in condensed matter). For continuous models this number is still well defined but not relevant for the bulk-edge correspondence: first it depends on the choice of boundary condition and furthermore, for a given boundary condition, this number can jump while varying continuously a parameter of the Hamiltonian (e.g. $\epsilon$), without closing the gap in the bulk, or even while varying $\omega$ with all parameters fixed. See Figure \ref{fig:edge_modes_largeepsilon}.

This paper shows that the correct edge number that matches in the bulk-edge correspondence is $n_\mathrm{top/bottom}\neq n_\mathrm{cross}(\omega) $. As discussed in Section \ref{sec:standardBEC} these quantities are the same only if $k_x$ and $H$ are bounded, which is not always true in continuous models. This allows for the existence of modes that leave a band without connecting another one, escaping in the infinite gaped region. This is reminiscent to the fact that the edge problem may not be compactified, even if the bulk problem is.

However we believe than $n_\mathrm{cross}(\omega)$ is still of interest because it counts a number of \emph{chiral} edge modes. Such modes are robust against defects on the boundary, as discussed in \cite{souslov2018topological}. In principle we expect these modes to be also stable under a disordered potential, so that  $n_\mathrm{cross}(\omega)$ must still be topological, but in a weaker sense that has to be investigated. We postpone the study of it to future work.\\

\paragraph{Coastal Kelvin are topologically protected in a weaker sense than equatorial Kelvin waves.} Coastal Kelvin waves are unidirectional edge states trapped along a boundary with impermeability condition ($v=0$ along the coast $y=0$), and with a trapping length scale given by then Rossby radius of deformation $L_d= c/f$ \cite{thomson1880}, with $c$ the phase speed of such waves. In Figure \ref{fig:edge_modes} of the shallow-water model they correspond to the edge mode with linear dispersion relation $\omega = c k_x$ with $c=1$ and $|k_x|<k_0$. We notice that this mode is always present in the spectrum, while the other edge modes depend on the boundary condition. We conjecture this to be true whenever the boundary condition includes the impermeability constraint. Moreover all additional edge modes have a trapping length scale that tends to zero as $\epsilon \rightarrow 0$, contrary to the coastal Kelvin wave that coincides in that limit with its analogue in absence of odd-viscosity ($k_0 \rightarrow \infty$). Finally it is robust to the continuous parameter deformation discussed above, and it is actually the only mode that is properly counted by spectral crossing $n_\mathrm{cross}(\omega)$. The coastal Kelvin wave seems therefore more robust than other edge modes in presence of a sharp boundary.

However, in the case without odd-viscosity, coastal Kelvin waves can be removed from the spectrum just by relaxing the impermeability constraint \cite{iga1995transition}, and we suspect the same to occur here, so that this mode is not topological in the strongest sense. This contrasts with unidirectional waves that are trapped along the equator of rotating atmospheres and oceans, and called equatorial Kelvin (and Yanai) waves by analogy. There the equator is an interface where $f$ changes sign. In contrast to a boundary, there is a canonical gluing condition for the interface, for which equatorial Kelvin wave is topological \cite{delplace2017topological,tauber2018odd}. This is due to the bulk-interface correspondence that does not suffer from any anomaly. This correspondence is a manifestation of Atiyah-Singer index theorem, that was noted in other physical problems and then generalized to a wider class of models \cite{faure2000topological,faure2001topological,fukui2012bulk,faure2019manifestation,bal2018continuous}. In the presence of a sharp boundary, as in this paper, the existence of an index theorem remains an open question.\\

\paragraph{Asymptotic edge modes are physical and can be detected.}
Finally we claim that the number of {``ghost''} edge modes at infinite frequency $n_+^\infty$ is not an abstract mathematical quantity but has physical consequences. First this number can be computed and even estimated numerically at finite $\omega$, as we did in Figures \ref{fig:scattering_top} and \ref{fig:dirac}. So the strict mathematical limit of infinite wavenumbers is not required to see this number: only a finite but sufficiently large spectral widow is required, for which the topological model is a valid description. Moreover the scattering principle described in Figure \ref{fig:intro_scattering}(a) probes the reflection of bulk waves transversely to the edge. So to speak, the bulk of the sample plays the role of a detector that probes what happens at the edge. In contrast to electrons in condensed matter, it is in principle possible in classical fluids and in optics to excite modes of the bulk with a specific frequency and wave-number. This could be implemented by measuring the reflection of these bulk excitations on a sample with a boundary. In the shallow-water model for equatorial waves, odd viscosity was considered only as a rather small regularizing parameter. But odd-viscous terms must actually be taken into account to properly describe active matter fluids and photonic systems where microscopic time reversibility is broken \cite{souslov2018topological,van2018photon}. In those cases, it should be possible to implement the measure of the scattering phase and thus detect the presence of asymptotic edge modes. 

\begin{acknowledgments}
C. T. is grateful to Gian Michele Graf and Hansueli Jud for many insightful discussions. P. D. and A.V. were partly funded by  ANR-18-CE30-0002-01 during this work.
\end{acknowledgments}

\newpage 

\appendix

\section{Allowed boundary conditions \label{app:self-adjoint}}

The allowed boundary conditions are constrained by looking at the self-adjointness of the problem. Rewriting \eqref{eq:general_model} as $\ii \partial_t \psi = H \psi$ with $\psi = (\eta,u,v)$ we impose the condition $\langle \phi, H\psi \rangle= \langle H \phi, \psi \rangle$, for any $\phi,\psi \in L^2(\mathbb R \times \mathbb R^+)$. After a few integration by parts we end up with
\begin{align}
\label{eq:self-adjoint-condition}
\ii \int_{\mathbb R} \dd x \Big( & v_1^* (\eta_2 + \epsilon \partial_y u_2) + (\eta_1^* + \epsilon \partial_y u_1^*) v_2  \cr 
& -\epsilon\big((\partial_y v_1^*) u_2  + u_1^* \partial_y v_2\big)\Big) \Big|_{y=0} = 0
\end{align}
which restricts the possible boundary conditions at $y=0$. We deduce that in general, only two constraints are required on $(\eta,u,v)$. In particular \eqref{eq:boundary_condition_DD}, \eqref{eq:boundary_condition_DM} and \eqref{eq:boundary_condition_DS} are solution to \eqref{eq:self-adjoint-condition}, but there exists many other possibilities for the shallow-water model. 

Similarly, the Hermitian boundary conditions at $y=0$ for the regularized Dirac model \eqref{eq:Dirac_regularized} have to satisfy
\begin{align}
\int_\mathbb{R}\dd x \Big(-\phi_1^*\chi_2 +  \phi_2^* \chi_1 
& -\epsilon(\phi_1^* \partial_y\chi_1 - \chi_1\partial_y\phi_1^*   \cr -\phi_2^*\partial_y \chi_2 &+ \chi_2\partial_y\phi_2^*)\Big)\Big|_{y=0}=0    \, .
\end{align} 
This is the case for the two boundary conditions \eqref{eq:boundary_condition_Dirac1} and \eqref{eq:boundary_condition_Dirac2} discussed in the main text.

\section{Solving the edge problem \label{app:edge_problem}}

In this appendix we solve the system of ODE \eqref{eq:edge_ode_v} and \eqref{eq:edge_ode_u} in $u$ and $v$ (we dropped the hat to simplify the notations).  In the gaped region of the $(k_x,\omega)$-parameter plane, we look for solutions that vanish as $y \rightarrow \infty$. First we compute all such modes that could exist in general, and then specify each boundary condition and see the compatible solutions that persist. Moreover in the following we assume $f>0$, $\epsilon >0$ and $f \epsilon< 1/4$. For the general problem we proceed by disjunction. First note that $u\equiv 0$ leads to $v\equiv 0$, so this case is trivial.

\paragraph{Case 1: $v\equiv0$}

From \eqref{eq:edge_ode_v} we infer $\omega^2=k_x^2$ that has two branches, $\omega = \pm k_x$. 
 Since $v\equiv0$ the solution to \eqref{eq:edge_ode_u} is of the form
\begin{equation}
u(y) = A \ee^{q_{+} y} + B \ee^{q_{-} y}
\end{equation}
where
\begin{equation}\label{eq:q_up}
q_{\pm} = - \dfrac{1}{2\epsilon} \big(\dfrac{k_x}{\omega} \pm \sqrt{1+4\epsilon(\epsilon k_x^2-f)}\big)
\end{equation}
that is always well defined as long as $f\epsilon \leq 1/4$. Notice that $q_\pm$, $A$ and $B$ depend on $k_x$ and $\omega$. For $u$ to vanish at $y\rightarrow \infty$ we have either $q_{\pm} <0$ or $A/B =0$.

\paragraph*{Case 1.a: $\omega=k_x$} One has $q_{+}<0$ for all $k_x$ and   $q_{-}<0$ only for $|k_x| < k_0 := \sqrt{f/\epsilon}$, so that 
\begin{equation}
u(y) = \left\lbrace \begin{array}{ll}
A \ee^{q_{+} y} + B \ee^{q_{-} y}, & |k_x| < k_0,\\
A \ee^{q_{+} y}, & |k_x| \geq k_0.
\end{array}\right.
\end{equation}

\paragraph*{Case 1.b: $\omega=-k_x$} One has $q_->0$ for all $k_x$ and $q_{+}<0$ only for $|k_x| > k_0$, so that 
\begin{equation}
u(y) = \left\lbrace \begin{array}{ll}
0 & |k_x| \leq k_0,\\
A \ee^{q_{+} y}, & |k_x| > k_0.
\end{array}\right.
\end{equation}

\paragraph{Case 2: $v \neq 0$ and $\omega^2=k_x^2$}

We first solve \eqref{eq:edge_ode_v} that is homogeneous for $v$. The solution is of the form
\begin{equation}
v(y) = A \ee^{r_+ y} + B \ee^{r_- y}
\end{equation}
with 
\begin{equation}\label{eq:r_up}
r_{\pm} =  \dfrac{1}{2\epsilon} \big(\dfrac{k_x}{\omega} \pm \sqrt{1+4\epsilon(\epsilon k_x^2-f)}\big)
\end{equation}
The solution of \eqref{eq:edge_ode_u} is a superposition of a homogeneous part, already given in the previous section, and a particular solution depending on the solution for $v$. Namely
\begin{equation}
u(y) =  C \ee^{q_{+} y} + D \ee^{q_{-} y} + \alpha_+ A \ee^{r_+ y} + \alpha_{-} B \ee^{r_- y}
\end{equation}
with $q_\pm$ given in \eqref{eq:q_up} and 
\begin{equation}
\alpha_{\pm} = - \dfrac{i}{\omega}(r_{\pm}^2 + \omega^2) \big(\epsilon r_{\pm}^2 + \dfrac{k_x}{\omega}r_{\pm} + f -\epsilon k_x^2\big)^{-1}
\end{equation}
\paragraph*{Case 2.a: $\omega=k_x$} One has $r_{+}\geq0$ for all $k_x$ and $r_{-}\geq0$ for $|k_x|\leq k_0$ so that
\begin{equation}
v(y) = \left\lbrace \begin{array}{ll}
0 & |k_x| \leq k_0,\\
B \ee^{r_{-} y}, & |k_x| > k_0.
\end{array}\right.
\end{equation}
Note that for $|k_x| \leq k_0$ we are back to Case 1, so we have to omit this region here to avoid double counting. Consequently
\begin{equation}
u(y) = C \ee^{q_+y} + \alpha_{-} B \ee^{r_{-} y}, \qquad |k_x| > k_0.
\end{equation}
\paragraph*{Case 2.b: $\omega=-k_x$}One has $r_{-} <0$ for all $k_x$ and $r_{+}<0$ for $|k_x|<k_0$, so that
\begin{equation}
v(y) = \left\lbrace \begin{array}{ll}
A \ee^{r_+ y} + B \ee^{r_- y} & |k_x| \leq k_0,\\
B \ee^{r_- y}, & |k_x| > k_0.
\end{array}\right.
\end{equation}
and
\begin{equation}
u(y) = \left\lbrace \begin{array}{ll}
\alpha_+ A \ee^{r_+ y} + \alpha_{-} B \ee^{r_- y} & |k_x| \leq k_0,\\
C \ee^{q_+y} + \alpha_- B \ee^{r_- y}, & |k_x| > k_0.
\end{array}\right.
\end{equation}

\paragraph{Case 3: $v \neq 0$ and $\omega^2\neq k_x^2$}

In that case $u$ is entirely fixed by $v$ through equation \eqref{eq:edge_ode_v}, and one can moreover combine \eqref{eq:edge_ode_v} and \eqref{eq:edge_ode_u} to get a fourth order homogeneous equation for $v$:
\begin{equation}\label{eq:yanai_v_order4}
\Big(\epsilon^2 \partial^{(4)}_{y}  + (2\epsilon( f-\epsilon k_x^2)-1) \partial^{(2)}_{y} + ( f-\epsilon k_x^2)^2-(\omega^2-k_x^2)\Big)v = 0
\end{equation} 
The corresponding algebraic equation always admits real solutions as long as $f \epsilon \leq 1/4$, given by $s^2 = S_\pm$ with
\begin{equation}\label{eq:def_Spm}
S_\pm = \dfrac{1}{2\epsilon^2} \Big( 1+2\epsilon(\epsilon k_x^2- f) \pm \sqrt{1+4 \epsilon (\epsilon \omega^2-f)} \Big)
\end{equation}
In the gapped region, one has
\begin{equation}\label{eq:interface_bulk_limit}
k_x^2-\omega^2 + (f-\epsilon k_x^2)^2 > 0
\end{equation}
leading to four real solutions to \eqref{eq:yanai_v_order4}
\begin{equation}
s_{1} = \sqrt{S_+}, \quad s_{2} = \sqrt{S_-}, \quad s_{3} = -\sqrt{S_+}, \quad s_{4} = -\sqrt{S_-}
\end{equation}
Notice that by construction $s_{1/2} >0$ and $s_{3/4} <0$ regardless of $k_x, \omega$ or $f$. 
Consequently,
\begin{equation}\label{general_yanai_v}
v(y) = V_3 \ee^{s_3 y} + V_4 \ee^{s_4 y}
\end{equation}
and by \eqref{eq:edge_ode_v}
\begin{equation}
u(y) =  \lambda_3 V_3 \ee^{s_3 y} + \lambda_4 V_4 \ee^{s_4 y}
\end{equation}
where 
\begin{equation}\label{eq:def_lambda}
\lambda_{i} = \dfrac{\omega}{\ii (\omega^2-k_x^2)} \big( \epsilon s_{ i}^2 - \dfrac{k_x}{\omega}s_{ i} + f - \epsilon k_x^2 \big).
\end{equation}

\subsection{Edge modes}

Now we specify a boundary condition from \eqref{eq:boundary_conditions} and look at the modes from the previous section that are compatible with it.

\paragraph{Dirichlet/Dirichlet (DD)}

Here we impose \eqref{eq:boundary_condition_DD}, namely $u=v=0$ at $y=0$. In Case 1, we infer immediately
\begin{equation}
u(y) = \left\lbrace \begin{array}{ll}
A (\ee^{q_{+} y} -  \ee^{q_{-} y}), & |k_x| < k_0, \quad \mathrm{and} \quad \omega=k_x\\
0 & \mathrm{otherwise.}
\end{array}\right.
\end{equation}
and $v\equiv 0$. One has one mode (\textit{i.e} one free parameter $A$) living in a compact region (see Figure \ref{fig:edge_modes}(a)). In Case 2 the solutions are trivial for $\omega=k_x$, and $\omega =-k_x$ for $|k_x| > k_0$. The last possibility is
\begin{equation}
v(y) = A (\ee^{r_+ y} - \ee^{r_- y}), \qquad |k_x| \leq k_0,
\end{equation}
but only if $u(y) = A (\alpha_+\ee^{r_+ y} - \alpha_-\ee^{r_- y})$ vanishes at $y=0$, which implies that $\alpha_+(k_x,-k_x)-\alpha_-(k_x,-k_x) = 0$. This generically occurs only for a finite number of $k_x$ points, that are actually part of the edge modes from Case 3. Apart from that there is no mode in that case. In Case 3 the region of compatibility with the boundary conditions is given by 
\begin{equation}\label{d/d_yanai}
0 = \det \begin{pmatrix}
1 & 1 \\ \lambda_3(k_x,\omega) & \lambda_4(k_x,\omega) 
\end{pmatrix}=\lambda_4(k_x,\omega) - \lambda_3(k_x,\omega)
\end{equation}
for $(k_x,\omega)$ in the gapped region but away from the branches $k_x^2 =\omega^2$ that are forbidden by assumption. The latter constraint leads to
\begin{equation}
\epsilon (s_3(k_x,\omega) + s_4(k_x,\omega)) - \frac{k_x}{\omega} =0
\end{equation}
that is plotted in Figure \ref{fig:edge_modes}(a). We have one mode in each gap that stops in a bulk band at $k_x=k_1$ with
\begin{equation}\label{defk1}
    k_1 := \pm k_0 \sqrt{1- \dfrac{3}{4 f \epsilon} \left(1-\sqrt{1-\dfrac{16}{9} f\epsilon} \right)}
\end{equation}
on one side and saturates at $\pm \tfrac{1}{2 \epsilon}$ as $k_x \rightarrow \pm \infty$. Along this curve, the kernel of the matrix appearing in \eqref{d/d_yanai} is generated by $(1, -1)$, so that \eqref{general_yanai_v} is a solution for $V_4 = -V_3$, namely
\begin{equation}
v(y) = V_3(\ee^{s_3 y} - \ee^{s_4 y} ), \quad \qquad u(y) = \lambda_3 V_3(\ee^{s_3 y} - \ee^{s_4 y} ).
\end{equation}
Thus we have one edge mode in that case.

\paragraph{Dirichlet/Membrane (DM)}

Here we impose \eqref{eq:boundary_condition_DM}, namely $v=0$ and $\partial_x u + \partial_y v =0$ at $y=0$. For the normal modes the latter condition can be rewritten $-\ii k_x u + \partial_y v =0$. In Case 1 where $v\equiv0$ it is equivalent to $u=0$ at the boundary, so this is similar to the Dirichlet/Dirichlet problem from the previous section. Hence we have one mode given by
\begin{equation}
u(y) = \left\lbrace \begin{array}{ll}
A (\ee^{q_{+} y} -  \ee^{q_{-} y}), & |k_x| < k_0, \quad \mathrm{and} \quad \omega=k_x\\
0 & \mathrm{otherwise.}
\end{array}\right.
\end{equation}
For Case 2.a, the solutions are trivial due to $v(0)=0$. For Case 2.b, where $\omega=-k_x$ this condition implies
\begin{equation}
v(y) = \left\lbrace \begin{array}{ll}
A (\ee^{r_+ y} -  \ee^{r_- y}) & |k_x| \leq k_0,\\
0 & |k_x| > k_0.
\end{array}\right.
\end{equation}
and
\begin{equation}
u(y) = \left\lbrace \begin{array}{ll}
A(\alpha_+  \ee^{r_+ y} - \alpha_{-}  \ee^{r_- y}) & |k_x| \leq k_0,\\
C \ee^{q_+y}, & |k_x| > k_0.
\end{array}\right.
\end{equation}
For $|k_x|>k_0$, the boundary condition implies $C=0$ and for $|k_x|\leq k_0$ there exists a non-trivial solution only if 
\begin{align}\label{di_extra}
-\ii k_x  \big( & \alpha_+(k_x,-k_x) -\alpha_-(k_x,-k_x) \big) \cr & + r_+(k_x,-k_x) - r_-(k_x,-k_x) = 0
\end{align}
One can check (e.g. numerically) that this equation is never satisfied for $k_x \in \mathbb R$. Finally for Case 3, $v(0)=0$ implies $V_4=-V_3$ and the membrane condition leads to  \begin{equation}- \ii k_x (\lambda_3-\lambda_4) + (s_3-s_4)=0  \label{eq:membrane_lambda_s}
\end{equation}
that simplifies to
\begin{equation}
\dfrac{\epsilon k_x (s_3(k_x,\omega) + s_4(k_x,\omega)) - \omega}{\omega^2-k_x^2} = 0.
\end{equation}
One can check numerically that no edge mode appears in that case. In conclusion we only have one edge mode, as illustrated in Figure \ref{fig:edge_modes}(b).

\paragraph{Dirichlet/Stress-free (DS)}

Here we impose \eqref{eq:boundary_condition_DS}, namely $v=0$ and $\partial_x u-\partial_y v=$ at $y=0$. Up to a change of sign we can solve this problem based on the derivation for condition DM from the previous section. The result is plotted in Figure \ref{fig:edge_modes}(c). Case 1 is unchanged since $v\equiv 0$ and we have the usual Kelvin wave. Case 2 has non-trivial solution for $\omega=-k_x$ only if 
\begin{align}\label{ds_extra}
-\ii k_x (&\alpha_+(k_x,-k_x) -\alpha_-(k_x,-k_x) )\cr & - r_+(k_x,-k_x) + r_-(k_x,-k_x) = 0
\end{align}
that vanishes for two values of $k_x$, which are actually part of the solution of Case 3. Case 3 reduces to 
\begin{equation}
\dfrac{\epsilon k_x \omega (s_3(k_x,\omega) + s_4(k_x,\omega)) + \omega^2-2k_x^2}{\omega^2-k_x^2} = 0.
\end{equation}
It has a non-trivial solution with three branches: two similar to the Dirichlet/Dirichlet (no-slip) boundary condition, but that saturates at $\omega = \mp \tfrac{1}{\epsilon}$ when $k_x \rightarrow \pm \infty$. These branches stop in the bulk bands at $k_x = \pm k_2$ and the apparent discontinuity in Figure \ref{fig:edge_modes}(c) is only an artifact, cured by the two points from Case 2 (see inset of Figure \ref{fig:edge_modes}(c)). Finally the third branch looks like $\omega = 2k_x$ near $k_x=0$ and stops at $k_x=\pm k_3$ when entering the bulk bands. There are no simple explicit expressions for $k_2$ and $k_3$ (in contrast to \eqref{defk1}), but they can be anyway estimated numerically with arbitrary precision.

\section{Scattering data \label{app:scat_data}}

The scattering matrix is obtained by requiring a boundary condition on the scattering state \eqref{eq:psi_scat_bottom} that is a superposition the bulk normal mode $\hat \psi_0$ (or $\hat \psi_\infty$) for different values of $\kappa$.

\subsection{Bottom of the band}

For the bottom of the band $\omega_+$ we use $\hat \psi_0:= (\eta^0,u^0,v^0)$ (we drop the hat to simplify the notation). In the following we denote $u^0_\mathrm{in} := u^0(k_x,-\kappa)$, $u^0_\mathrm{out} := u^0(k_x,\kappa)$ and $\tilde u^0 = u^0(k_x,\tilde \kappa_-(k_x,\kappa))$, where $\tilde \kappa_-(k_x,\kappa) = - \ii \sqrt{K_-}$ (see \eqref{eq:Kpm} and above), and similarly for $v^0$. The explicit expressions for $u^0$ and $v^0$ are given in \eqref{eq:defPsi+} up to a phase multiplication by $\lambda_0$.

\paragraph{Dirichlet/Dirichlet (DD)} Here we impose \eqref{eq:boundary_condition_DD}, namely $u=v=0$ at $y=0$. From \eqref{eq:psi_scat_bottom} we infer 
\begin{align}
&\alpha u_\mathrm{in}^0 + \beta u_\mathrm{out}^0 + \gamma \tilde u^0  =0\cr
&\alpha v_\mathrm{in}^0 + \beta v_\mathrm{out}^0 + \gamma \tilde v^0  =0
\end{align}
so that
\begin{equation}\label{eq:S_DD}
S(k_x,\kappa) = \dfrac{v_\mathrm{in}^0 \tilde u^0 - u_\mathrm{in}^0 \tilde v^0}{ u_\mathrm{out}^0 \tilde v^0- v_\mathrm{out}^0 \tilde u^0}
\end{equation}
The argument of $S$ is plotted in Figure \ref{fig:scattering_bottom}(a) with respect to $k_x$ and for several small values of $\kappa$.

\paragraph{Dirichlet/Membrane (DM)} Here we impose \eqref{eq:boundary_condition_DM}, namely $v=0$ and $\partial_x u + \partial_y v =0$ at $y=0$. From \eqref{eq:psi_scat_bottom} we infer 
\begin{align}
&\alpha \big( k_x u^0_\mathrm{in} -  \kappa v^0_\mathrm{in}\big) + \beta ( k_x u^0_\mathrm{out} + \kappa v^0_\mathrm{out} ) \cr
& \hspace{3cm}  + \gamma ( k_x \tilde u^0 + \tilde \kappa_- \tilde v^0 ) =0\cr
&\alpha v^0_\mathrm{in} + \beta v^0_\mathrm{out} + \gamma \tilde v^0  =0
\end{align}
so that
\begin{equation}\label{eq:S_DM}
S(k_x,\kappa) = -\dfrac{(k_x u^0_\mathrm{in} - \kappa v^0_\mathrm{in})\tilde v^0 - v^0_\mathrm{in}(k_x \tilde u^0 + \tilde \kappa_- \tilde v^0)}{(k_x u^0_\mathrm{out}+\kappa v^0_\mathrm{out}) \tilde v^0 - v^0_\mathrm{out}(k_x \tilde u^0 + \tilde \kappa_- \tilde v^0)}
\end{equation}
The argument of $S$ is plotted in Figure \ref{fig:scattering_bottom}(b) with respect to $k_x$ and for several small values of $\kappa$.

\paragraph{Dirichlet/Stress-free (DS)} Here we impose \eqref{eq:boundary_condition_DS}, namely $v=0$ and $\partial_x u-\partial_y v=$ at $y=0$. From \eqref{eq:psi_scat_bottom} we infer
\begin{align}
&\alpha \big( k_x u^0_\mathrm{in} +  \kappa v^0_\mathrm{in}\big) + \beta ( k_x u^0_\mathrm{out} - \kappa v^0_\mathrm{out} )\cr
& \hspace{3cm}  + \gamma ( k_x \tilde u^0 - \tilde \kappa_- \tilde v^0 ) =0\cr
&\alpha v^0_\mathrm{in} + \beta v^0_\mathrm{out} + \gamma \tilde v^0  =0
\end{align}
so that
\begin{equation}\label{eq:S_DS}
S(k_x,\kappa) = -\dfrac{(k_x u^0_\mathrm{in} + \kappa v^0_\mathrm{in})\tilde v^0 - v^0_\mathrm{in}(k_x \tilde u^0 - \tilde \kappa_- \tilde v^0)}{(k_x u^0_\mathrm{out}-\kappa v^0_\mathrm{out}) \tilde v^0 - v^0_\mathrm{out}(k_x \tilde u^0 - \tilde \kappa_- \tilde v^0)}
\end{equation}
The argument of $S$ is plotted in Figure \ref{fig:scattering_bottom}(c) with respect to $k_x$ and for several small values of $\kappa$.

\subsection{Scattering at infinity}

To explore the infinite upper limit of the band $\omega_+$ the scattering state $\eqref{eq:psi_scat_bottom}$ is computed using $\hat \psi_\infty$ instead of $\hat \psi_0$ but the derivation of $S$ is formally the same than in the previous section. Thus the expression of $S$ in that case is given by \eqref{eq:S_DD}, \eqref{eq:S_DM} or \eqref{eq:S_DS} (respectively for DD, DM and DS) where we replace $u_\mathrm{in}^0$, $u_\mathrm{out}^0$ and $\tilde u^0$ by $u_\mathrm{in}^\infty$, $u_\mathrm{out}^\infty$ and $\tilde u^\infty$, and similarly for $v$. The explicit expressions of theses quantities come from \eqref{eq:defPsi+} up to a phase multiplication by $\lambda_\infty$. The argument of $S$ is plotted for each boundary condition in Figure \ref{fig:scattering_top} with respect to $k_x$ and for several large values of $\kappa$.

\section{Regularized Dirac Hamiltonian}
\label{sec:dirac}
\subsection{Edge modes}
We aim at calculating the edge modes for a semi-infinite plane $(x,y)\in \mathbb{R}\times [0,\infty]$ geometry with two different boundary conditions A and B defined in equations \eqref{eq:boundary_condition_Dirac1} and \eqref{eq:boundary_condition_Dirac2}. Following the same lines as for the shallow-water model, boundary modes  are obtained as the linear combination  
\begin{equation}
\Psi=\begin{pmatrix}
A_- \\ B_- 
\end{pmatrix}
\ee^{-K_- y}
+\begin{pmatrix}
A_+ \\ B_+ 
\end{pmatrix}
\ee^{-K_+ y}
\label{eq:sup_ev}
\end{equation}
where the evanescent modes
\begin{equation}
\begin{pmatrix}
A_\pm \\ B_\pm 
\end{pmatrix}
\ee^{-K_\pm y}
\end{equation}
are solutions of {$H_{\text{half plane}} \Psi = E \Psi$ with} 
\begin{equation}
    H_{\text{half plane}} = \begin{pmatrix}
    m +\epsilon k_x^2-\epsilon \partial_y^2 & k_x +\partial_y \\
    k_x - \partial_y & -m -\epsilon k_x^2 + \epsilon \partial_y^2
    \end{pmatrix} \, .
\end{equation} 
A direct calculation leads to 
\begin{equation}
    K_\pm=\frac{1}{\sqrt{2}\epsilon} \sqrt{2\epsilon(m-\epsilon k_x^2)+1\pm \sqrt{1-4 \epsilon(m-\epsilon E^2)}}\, .
\end{equation}
Notice that, for simplicity, we have only considered the case where $\sqrt{1-4 \epsilon(m-\epsilon E^2)}$ is real in the decomposition \eqref{eq:sup_ev}, that is satisfied when $4 |\epsilon m| <1$. Then, defining $\lambda_\pm$ as $B_\pm = \lambda_\pm A_\pm$, one gets 
\begin{equation}
\label{eq:lambda}
    \lambda_\pm(E,k_x) = \frac{k_x+K_\pm}{m-\epsilon k_x^2+\epsilon K_\pm^2+E}\, .
\end{equation}

Finally, inserting \eqref{eq:sup_ev} with \eqref{eq:lambda} into the boundary conditions A and B respectively yields
\begin{subequations}\label{eq:edge_mode_Dirac}
\begin{align}
\label{eq:edge_mode_Dirac1}\text{A:}& \quad \lambda_+(E,k_x) - \lambda_-(E,k_x) =0 \\
\label{eq:edge_mode_Dirac2}
\text{B:}&   \quad 
\lambda_+(E,k_x)(K_+-k_x) - \lambda_-(E,k_x)(K_--k_x) =0 \ .
\end{align}
\end{subequations}
These two implicit equations over $E$ and $k_x$ give the dispersion relation of the evanescent modes compatible with the corresponding boundary conditions A and B. These dispersion relations are plotted in Figure \ref{fig:dirac}.

\subsection{Chern number}
When the mass $m$ is fixed, the regularization $\epsilon\neq0$, allows a well defined (integer-valued)
 first Chern number
\begin{equation}
C_\pm= \frac{\ii}{2\pi} \int_{\mathbb{R}^2} \dd k_x \dd k_y  \left( \braket{\partial_{k_x}\psi_\pm|\partial_{k_y}\psi_\pm}-\braket{\partial_{k_y}\psi_\pm|\partial_{k_x}\psi_\pm}\right)
\end{equation}
for each bulk eigenstate $\psi_\pm(k_x,k_y)$ of energy $E_\pm(k_x,k_y)=\pm\sqrt{k^2+(m+\epsilon k^2)^2}$, solutions of \eqref{eq:Dirac_bulk}
There are several ways to compute the Chern number. One of them consists in noticing that it coincides with the degree of the map from $S^2$ (the compactified $\mathbb{R}^2$ plane) to $S^2$ (the projective space for normalized spinors) \cite{volovik1988analogue}. An alternative way, that is also convenient to compute the scattering states in the following, consists in looking for the phase singularities of the normalized eigenstates $\psi_\pm (k_x,k_y)$.
indeed $\psi_\pm (k_x,k_y)$
may have a phase singularity at $k\sim 0$ and/or at $k \sim \infty$ that can be  cured locally by a gauge choice of the phase, but not necessarily removed for any point of the plane $(k_x,k_y)$. This is a topological property of the model that is captured by the first Chern number. 

In particular, the behaviour of the eigenstate of positive energy
\begin{equation}
    \hat{\psi}_+ = \frac{1}{\sqrt{2}\sqrt{E_+^2-E_+(m+\epsilon k^2)}}
    \begin{pmatrix}
    k_x - \ii k_y\\
    E_+-m-\epsilon k^2
    \end{pmatrix}
    \label{eq:psi+}
\end{equation}
depends on the sign of the mass term as
\begin{align}
\label{eq:psi0}
\hat{\psi}_+ \underset{0}{\sim} \left\{
\begin{array}{ll}
\begin{pmatrix}
0\\ 1
\end{pmatrix} 
\quad &\text{for}\ m<0\\ 
\begin{pmatrix}
\ee^{-\ii \phi}\\ 0
\end{pmatrix} 
\quad &\text{for}\ m>0 \, .
\end{array}
\right. 
\end{align}
It is regular at $k\sim 0$ when $m <0$, but has a phase singularity when $m>0$.
This phase singularity can be removed by the gauge transformation $\psi_0 = \lambda \hat{\psi}_+$ with $\lambda=\ee^{\ii \phi}=k^{-1}(k_x+\ii k_y)$.  Similarly 
\begin{align}
\label{eq:psiinfty}
\hat{\psi}_+ \underset{\infty}{\sim} \left\{
\begin{array}{ll}
\begin{pmatrix}
\ee^{-\ii \phi}\\ 0
\end{pmatrix} 
\quad &\text{for}\ \epsilon<0\\ 
\begin{pmatrix}
0\\ 1
\end{pmatrix} 
\quad &\text{for}\ \epsilon>0
\end{array}
\right.
\end{align}
so that $\hat{\psi}_+$ is regular at $k\sim \infty$ when $\epsilon <0$ but has the same phase singularity as at $k\sim 0$ for $\epsilon>0$. This singularity is thus removed from $k\sim \infty$ with the same gauge transformation.  
The Chern number captures the impossibility to remove the phase singularity at both $k\sim 0$ and $k\sim \infty$ by the a global choice of phase. Thus, it follows from \eqref{eq:psi0} and \eqref{eq:psiinfty} that the Chern number of the positive energy band vanishes when $\text{sgn}(m) =- \text{sgn}(\epsilon)$. Finally, a direct calculation leads to 
\begin{equation}
    C_\pm = \pm\frac{\text{sign}(m)+\text{sign}(\epsilon)}{2} 
\end{equation}
that only takes integer values. In particular, one recovers the so-called ``half-Chern number'' for the usual (un-regularized) massive two-dimensional Dirac equation when $\epsilon=0$.

\subsection{Scattering matrices}
For each boundary conditions \eqref{eq:boundary_condition_Dirac1} and \eqref{eq:boundary_condition_Dirac2}, the scattering matrix  \eqref{eq:scattering_matrix} is obtained at $\kappa\sim0/\infty$ from the scattering state \eqref{eq:scattering}, by taking a local regular section, i.e. by choosing a local gauge such that the bulk eigenstate is singled-valued at $k\sim0/\infty$. Focusing on $\psi_+$, \eqref{eq:psi0} and \eqref{eq:psiinfty} indicate that \eqref{eq:psi+} can be used to construct the scattering state around $k\sim 0$ when $m<0$ and at $k\sim \infty$ when $\epsilon >0$, while one must use $\lambda \hat{\psi}_\pm$ otherwise.

Denoting $\psi_+^{i}= (\phi_1^{i} , \phi_2^{i})^T$, a smooth section of $\psi_+$ at $k\sim i= \{0,\infty\}$, the scattering matrices $S_i(k_x,\kappa)$ at $k\sim i$ for the band of positive energy are found to be
\begin{subequations}\label{eq:scattering_Dirac}
\begin{align}
\label{eq:scattering_Dirac1}\text{A:}& \quad S_i(k_x,\kappa) = \frac{\phi^i_{1,\text{in}}\phi^i_{2,\text{b}} -\phi^i_{2,\text{in}} \phi^i_{1,\text{b}}}{\phi^i_{2,\text{out}} \phi^i_{1,\text{b}} - \phi^i_{1,\text{out}} \phi^i_{2,\text{b}}} \\
\label{eq:scattering_Dirac2}
\text{B:}&   \quad S_i(k_x,\kappa) = \notag \\ 
&-\frac{
(k_x- |\tilde{\kappa}_-|)\phi^i_{2,\text{b}}\phi^i_{1_\text{in}}-
(\ii \kappa +k_x)\phi^i_{1,\text{b}}\phi^i_{2_\text{in}}}
{(k_x- |\tilde{\kappa}_-|)\phi^i_{2,\text{b}}\phi^i_{1_\text{out}}-
(-\ii \kappa +k_x)\phi^i_{1,\text{b}}\phi^i_{2_\text{out}}}
\end{align}
\end{subequations}
for the boundary conditions A and B, and where
\begin{equation}
    \tilde{\kappa}_-(k_x,\kappa) = -\ii \sqrt{\kappa^2 + \frac{1-2\epsilon(m-\epsilon k_x^2)}{\epsilon^2}}
\end{equation}
Their argument is ploted as a function of $k_x$ for different values of $\kappa$ in Figure \ref{fig:dirac}. Its winding gives, in unit of $2\pi$, the number of boundary states that enter the positive energy band by below (at finite $k$) or from the top (at $k\sim \infty$), so that the bulk-boundary correspondence is satisfied.

\bibliographystyle{apsrev4-1}
\bibliography{anomalous_bulkedge}

\begin{thebibliography}{57}%
\makeatletter
\providecommand \@ifxundefined [1]{%
 \@ifx{#1\undefined}
}%
\providecommand \@ifnum [1]{%
 \ifnum #1\expandafter \@firstoftwo
 \else \expandafter \@secondoftwo
 \fi
}%
\providecommand \@ifx [1]{%
 \ifx #1\expandafter \@firstoftwo
 \else \expandafter \@secondoftwo
 \fi
}%
\providecommand \natexlab [1]{#1}%
\providecommand \enquote  [1]{``#1''}%
\providecommand \bibnamefont  [1]{#1}%
\providecommand \bibfnamefont [1]{#1}%
\providecommand \citenamefont [1]{#1}%
\providecommand \href@noop [0]{\@secondoftwo}%
\providecommand \href [0]{\begingroup \@sanitize@url \@href}%
\providecommand \@href[1]{\@@startlink{#1}\@@href}%
\providecommand \@@href[1]{\endgroup#1\@@endlink}%
\providecommand \@sanitize@url [0]{\catcode `\\12\catcode `\$12\catcode
  `\&12\catcode `\#12\catcode `\^12\catcode `\_12\catcode `\%12\relax}%
\providecommand \@@startlink[1]{}%
\providecommand \@@endlink[0]{}%
\providecommand \url  [0]{\begingroup\@sanitize@url \@url }%
\providecommand \@url [1]{\endgroup\@href {#1}{\urlprefix }}%
\providecommand \urlprefix  [0]{URL }%
\providecommand \Eprint [0]{\href }%
\providecommand \doibase [0]{http://dx.doi.org/}%
\providecommand \selectlanguage [0]{\@gobble}%
\providecommand \bibinfo  [0]{\@secondoftwo}%
\providecommand \bibfield  [0]{\@secondoftwo}%
\providecommand \translation [1]{[#1]}%
\providecommand \BibitemOpen [0]{}%
\providecommand \bibitemStop [0]{}%
\providecommand \bibitemNoStop [0]{.\EOS\space}%
\providecommand \EOS [0]{\spacefactor3000\relax}%
\providecommand \BibitemShut  [1]{\csname bibitem#1\endcsname}%
\let\auto@bib@innerbib\@empty
\bibitem [{\citenamefont {Thouless}\ \emph {et~al.}(1982)\citenamefont
  {Thouless}, \citenamefont {Kohmoto}, \citenamefont {Nightingale},\ and\
  \citenamefont {den Nijs}}]{thouless1982quantized}%
  \BibitemOpen
  \bibfield  {author} {\bibinfo {author} {\bibfnamefont {D.~J.}\ \bibnamefont
  {Thouless}}, \bibinfo {author} {\bibfnamefont {M.}~\bibnamefont {Kohmoto}},
  \bibinfo {author} {\bibfnamefont {M.~P.}\ \bibnamefont {Nightingale}}, \ and\
  \bibinfo {author} {\bibfnamefont {M.}~\bibnamefont {den Nijs}},\ }\href@noop
  {} {\bibfield  {journal} {\bibinfo  {journal} {Physical Review Letters}\
  }\textbf {\bibinfo {volume} {49}},\ \bibinfo {pages} {405} (\bibinfo {year}
  {1982})}\BibitemShut {NoStop}%
\bibitem [{\citenamefont {Laughlin}(1981)}]{laughlin1981quantized}%
  \BibitemOpen
  \bibfield  {author} {\bibinfo {author} {\bibfnamefont {R.~B.}\ \bibnamefont
  {Laughlin}},\ }\href@noop {} {\bibfield  {journal} {\bibinfo  {journal}
  {Physical Review B}\ }\textbf {\bibinfo {volume} {23}},\ \bibinfo {pages}
  {5632} (\bibinfo {year} {1981})}\BibitemShut {NoStop}%
\bibitem [{\citenamefont {Halperin}(1982)}]{halperin1982quantized}%
  \BibitemOpen
  \bibfield  {author} {\bibinfo {author} {\bibfnamefont {B.~I.}\ \bibnamefont
  {Halperin}},\ }\href@noop {} {\bibfield  {journal} {\bibinfo  {journal}
  {Physical Review B}\ }\textbf {\bibinfo {volume} {25}},\ \bibinfo {pages}
  {2185} (\bibinfo {year} {1982})}\BibitemShut {NoStop}%
\bibitem [{\citenamefont {Hatsugai}(1993)}]{hatsugai1993chern}%
  \BibitemOpen
  \bibfield  {author} {\bibinfo {author} {\bibfnamefont {Y.}~\bibnamefont
  {Hatsugai}},\ }\href@noop {} {\bibfield  {journal} {\bibinfo  {journal}
  {Physical Review Letters}\ }\textbf {\bibinfo {volume} {71}},\ \bibinfo
  {pages} {3697} (\bibinfo {year} {1993})}\BibitemShut {NoStop}%
\bibitem [{\citenamefont {Hasan}\ and\ \citenamefont
  {Kane}(2010)}]{hasan2010colloquium}%
  \BibitemOpen
  \bibfield  {author} {\bibinfo {author} {\bibfnamefont {M.~Z.}\ \bibnamefont
  {Hasan}}\ and\ \bibinfo {author} {\bibfnamefont {C.~L.}\ \bibnamefont
  {Kane}},\ }\href@noop {} {\bibfield  {journal} {\bibinfo  {journal} {Reviews
  of Modern Physics}\ }\textbf {\bibinfo {volume} {82}},\ \bibinfo {pages}
  {3045} (\bibinfo {year} {2010})}\BibitemShut {NoStop}%
\bibitem [{\citenamefont {Hatsugai}(2009)}]{hatsugai2009bulk}%
  \BibitemOpen
  \bibfield  {author} {\bibinfo {author} {\bibfnamefont {Y.}~\bibnamefont
  {Hatsugai}},\ }\href@noop {} {\bibfield  {journal} {\bibinfo  {journal}
  {Solid State Communications}\ }\textbf {\bibinfo {volume} {149}},\ \bibinfo
  {pages} {1061} (\bibinfo {year} {2009})}\BibitemShut {NoStop}%
\bibitem [{\citenamefont {Isaev}\ \emph {et~al.}(2011)\citenamefont {Isaev},
  \citenamefont {Moon},\ and\ \citenamefont {Ortiz}}]{isaev2011bulk}%
  \BibitemOpen
  \bibfield  {author} {\bibinfo {author} {\bibfnamefont {L.}~\bibnamefont
  {Isaev}}, \bibinfo {author} {\bibfnamefont {Y.}~\bibnamefont {Moon}}, \ and\
  \bibinfo {author} {\bibfnamefont {G.}~\bibnamefont {Ortiz}},\ }\href@noop {}
  {\bibfield  {journal} {\bibinfo  {journal} {Physical Review B}\ }\textbf
  {\bibinfo {volume} {84}},\ \bibinfo {pages} {075444} (\bibinfo {year}
  {2011})}\BibitemShut {NoStop}%
\bibitem [{\citenamefont {Graf}\ and\ \citenamefont
  {Porta}(2013)}]{graf2013bulk}%
  \BibitemOpen
  \bibfield  {author} {\bibinfo {author} {\bibfnamefont {G.~M.}\ \bibnamefont
  {Graf}}\ and\ \bibinfo {author} {\bibfnamefont {M.}~\bibnamefont {Porta}},\
  }\href@noop {} {\bibfield  {journal} {\bibinfo  {journal} {Communications in
  Mathematical Physics}\ }\textbf {\bibinfo {volume} {324}},\ \bibinfo {pages}
  {851} (\bibinfo {year} {2013})}\BibitemShut {NoStop}%
\bibitem [{\citenamefont {Avila}\ \emph {et~al.}(2013)\citenamefont {Avila},
  \citenamefont {Schulz-Baldes},\ and\ \citenamefont
  {Villegas-Blas}}]{avila2013topological}%
  \BibitemOpen
  \bibfield  {author} {\bibinfo {author} {\bibfnamefont {J.~C.}\ \bibnamefont
  {Avila}}, \bibinfo {author} {\bibfnamefont {H.}~\bibnamefont
  {Schulz-Baldes}}, \ and\ \bibinfo {author} {\bibfnamefont {C.}~\bibnamefont
  {Villegas-Blas}},\ }\href@noop {} {\bibfield  {journal} {\bibinfo  {journal}
  {Mathematical Physics, Analysis and Geometry}\ }\textbf {\bibinfo {volume}
  {16}},\ \bibinfo {pages} {137} (\bibinfo {year} {2013})}\BibitemShut
  {NoStop}%
\bibitem [{\citenamefont {Schulz-Baldes}\ \emph {et~al.}(2000)\citenamefont
  {Schulz-Baldes}, \citenamefont {Kellendonk},\ and\ \citenamefont
  {Richter}}]{schulz2000simultaneous}%
  \BibitemOpen
  \bibfield  {author} {\bibinfo {author} {\bibfnamefont {H.}~\bibnamefont
  {Schulz-Baldes}}, \bibinfo {author} {\bibfnamefont {J.}~\bibnamefont
  {Kellendonk}}, \ and\ \bibinfo {author} {\bibfnamefont {T.}~\bibnamefont
  {Richter}},\ }\href@noop {} {\bibfield  {journal} {\bibinfo  {journal}
  {Journal of Physics A: Mathematical and General}\ }\textbf {\bibinfo {volume}
  {33}},\ \bibinfo {pages} {L27} (\bibinfo {year} {2000})}\BibitemShut
  {NoStop}%
\bibitem [{\citenamefont {Elgart}\ \emph {et~al.}(2005)\citenamefont {Elgart},
  \citenamefont {Graf},\ and\ \citenamefont {Schenker}}]{elgart2005equality}%
  \BibitemOpen
  \bibfield  {author} {\bibinfo {author} {\bibfnamefont {A.}~\bibnamefont
  {Elgart}}, \bibinfo {author} {\bibfnamefont {G.}~\bibnamefont {Graf}}, \ and\
  \bibinfo {author} {\bibfnamefont {J.}~\bibnamefont {Schenker}},\ }\href@noop
  {} {\bibfield  {journal} {\bibinfo  {journal} {Communications in mathematical
  physics}\ }\textbf {\bibinfo {volume} {259}},\ \bibinfo {pages} {185}
  (\bibinfo {year} {2005})}\BibitemShut {NoStop}%
\bibitem [{\citenamefont {Prodan}\ and\ \citenamefont
  {Schulz-Baldes}(2016)}]{prodan2016bulk}%
  \BibitemOpen
  \bibfield  {author} {\bibinfo {author} {\bibfnamefont {E.}~\bibnamefont
  {Prodan}}\ and\ \bibinfo {author} {\bibfnamefont {H.}~\bibnamefont
  {Schulz-Baldes}},\ }\href@noop {} {\emph {\bibinfo {title} {Bulk and boundary
  invariants for complex topological insulators}}}\ (\bibinfo  {publisher}
  {Mathematical Physics Studies, Springer},\ \bibinfo {year}
  {2016})\BibitemShut {NoStop}%
\bibitem [{\citenamefont {Graf}\ and\ \citenamefont
  {Shapiro}(2018)}]{Graf2018}%
  \BibitemOpen
  \bibfield  {author} {\bibinfo {author} {\bibfnamefont {G.~M.}\ \bibnamefont
  {Graf}}\ and\ \bibinfo {author} {\bibfnamefont {J.}~\bibnamefont {Shapiro}},\
  }\href {\doibase 10.1007/s00220-018-3247-0} {\bibfield  {journal} {\bibinfo
  {journal} {Communications in Mathematical Physics}\ }\textbf {\bibinfo
  {volume} {363}},\ \bibinfo {pages} {829} (\bibinfo {year}
  {2018})}\BibitemShut {NoStop}%
\bibitem [{\citenamefont {Rudner}\ \emph {et~al.}(2013)\citenamefont {Rudner},
  \citenamefont {Lindner}, \citenamefont {Berg},\ and\ \citenamefont
  {Levin}}]{rudner2013anomalous}%
  \BibitemOpen
  \bibfield  {author} {\bibinfo {author} {\bibfnamefont {M.~S.}\ \bibnamefont
  {Rudner}}, \bibinfo {author} {\bibfnamefont {N.~H.}\ \bibnamefont {Lindner}},
  \bibinfo {author} {\bibfnamefont {E.}~\bibnamefont {Berg}}, \ and\ \bibinfo
  {author} {\bibfnamefont {M.}~\bibnamefont {Levin}},\ }\href@noop {}
  {\bibfield  {journal} {\bibinfo  {journal} {Physical Review X}\ }\textbf
  {\bibinfo {volume} {3}},\ \bibinfo {pages} {031005} (\bibinfo {year}
  {2013})}\BibitemShut {NoStop}%
\bibitem [{\citenamefont {Asb{\'o}th}\ \emph {et~al.}(2014)\citenamefont
  {Asb{\'o}th}, \citenamefont {Tarasinski},\ and\ \citenamefont
  {Delplace}}]{asboth2014chiral}%
  \BibitemOpen
  \bibfield  {author} {\bibinfo {author} {\bibfnamefont {J.~K.}\ \bibnamefont
  {Asb{\'o}th}}, \bibinfo {author} {\bibfnamefont {B.}~\bibnamefont
  {Tarasinski}}, \ and\ \bibinfo {author} {\bibfnamefont {P.}~\bibnamefont
  {Delplace}},\ }\href@noop {} {\bibfield  {journal} {\bibinfo  {journal}
  {Physical Review B}\ }\textbf {\bibinfo {volume} {90}},\ \bibinfo {pages}
  {125143} (\bibinfo {year} {2014})}\BibitemShut {NoStop}%
\bibitem [{\citenamefont {Graf}\ and\ \citenamefont
  {Tauber}(2018)}]{graf2018bulk}%
  \BibitemOpen
  \bibfield  {author} {\bibinfo {author} {\bibfnamefont {G.~M.}\ \bibnamefont
  {Graf}}\ and\ \bibinfo {author} {\bibfnamefont {C.}~\bibnamefont {Tauber}},\
  }\href@noop {} {\bibfield  {journal} {\bibinfo  {journal} {Annales Henri
  Poincar{\'e}}\ }\textbf {\bibinfo {volume} {19}},\ \bibinfo {pages} {709}
  (\bibinfo {year} {2018})}\BibitemShut {NoStop}%
\bibitem [{\citenamefont {Shapiro}\ and\ \citenamefont
  {Tauber}(2018)}]{shapiro2018strongly}%
  \BibitemOpen
  \bibfield  {author} {\bibinfo {author} {\bibfnamefont {J.}~\bibnamefont
  {Shapiro}}\ and\ \bibinfo {author} {\bibfnamefont {C.}~\bibnamefont
  {Tauber}},\ }\href@noop {} {\bibfield  {journal} {\bibinfo  {journal} {arXiv
  preprint arXiv:1807.03251}\ } (\bibinfo {year} {2018})}\BibitemShut {NoStop}%
\bibitem [{\citenamefont {Volovik}(1988)}]{volovik1988analogue}%
  \BibitemOpen
  \bibfield  {author} {\bibinfo {author} {\bibfnamefont {G.}~\bibnamefont
  {Volovik}},\ }\href@noop {} {\bibfield  {journal} {\bibinfo  {journal}
  {Zhurnal Ehksperimental'noj i Teoreticheskoj Fiziki}\ }\textbf {\bibinfo
  {volume} {94}},\ \bibinfo {pages} {123} (\bibinfo {year} {1988})}\BibitemShut
  {NoStop}%
\bibitem [{\citenamefont {Raghu}\ and\ \citenamefont
  {Haldane}(2008)}]{raghu2008analogs}%
  \BibitemOpen
  \bibfield  {author} {\bibinfo {author} {\bibfnamefont {S.}~\bibnamefont
  {Raghu}}\ and\ \bibinfo {author} {\bibfnamefont {F.~D.~M.}\ \bibnamefont
  {Haldane}},\ }\href@noop {} {\bibfield  {journal} {\bibinfo  {journal}
  {Physical Review A}\ }\textbf {\bibinfo {volume} {78}},\ \bibinfo {pages}
  {033834} (\bibinfo {year} {2008})}\BibitemShut {NoStop}%
\bibitem [{\citenamefont {Rechtsman}\ \emph {et~al.}(2013)\citenamefont
  {Rechtsman}, \citenamefont {Zeuner}, \citenamefont {Plotnik}, \citenamefont
  {Lumer}, \citenamefont {Podolsky}, \citenamefont {Dreisow}, \citenamefont
  {Nolte}, \citenamefont {Segev},\ and\ \citenamefont
  {Szameit}}]{rechtsman2013photonic}%
  \BibitemOpen
  \bibfield  {author} {\bibinfo {author} {\bibfnamefont {M.~C.}\ \bibnamefont
  {Rechtsman}}, \bibinfo {author} {\bibfnamefont {J.~M.}\ \bibnamefont
  {Zeuner}}, \bibinfo {author} {\bibfnamefont {Y.}~\bibnamefont {Plotnik}},
  \bibinfo {author} {\bibfnamefont {Y.}~\bibnamefont {Lumer}}, \bibinfo
  {author} {\bibfnamefont {D.}~\bibnamefont {Podolsky}}, \bibinfo {author}
  {\bibfnamefont {F.}~\bibnamefont {Dreisow}}, \bibinfo {author} {\bibfnamefont
  {S.}~\bibnamefont {Nolte}}, \bibinfo {author} {\bibfnamefont
  {M.}~\bibnamefont {Segev}}, \ and\ \bibinfo {author} {\bibfnamefont
  {A.}~\bibnamefont {Szameit}},\ }\href@noop {} {\bibfield  {journal} {\bibinfo
   {journal} {Nature}\ }\textbf {\bibinfo {volume} {496}},\ \bibinfo {pages}
  {196} (\bibinfo {year} {2013})}\BibitemShut {NoStop}%
\bibitem [{\citenamefont {Lu}\ \emph {et~al.}(2014)\citenamefont {Lu},
  \citenamefont {Joannopoulos},\ and\ \citenamefont
  {Solja{\v{c}}i{\'c}}}]{lu2014topological}%
  \BibitemOpen
  \bibfield  {author} {\bibinfo {author} {\bibfnamefont {L.}~\bibnamefont
  {Lu}}, \bibinfo {author} {\bibfnamefont {J.~D.}\ \bibnamefont
  {Joannopoulos}}, \ and\ \bibinfo {author} {\bibfnamefont {M.}~\bibnamefont
  {Solja{\v{c}}i{\'c}}},\ }\href@noop {} {\bibfield  {journal} {\bibinfo
  {journal} {Nature Photonics}\ }\textbf {\bibinfo {volume} {8}},\ \bibinfo
  {pages} {821} (\bibinfo {year} {2014})}\BibitemShut {NoStop}%
\bibitem [{\citenamefont {Peano}\ \emph {et~al.}(2015)\citenamefont {Peano},
  \citenamefont {Brendel}, \citenamefont {Schmidt},\ and\ \citenamefont
  {Marquardt}}]{peano2015topological}%
  \BibitemOpen
  \bibfield  {author} {\bibinfo {author} {\bibfnamefont {V.}~\bibnamefont
  {Peano}}, \bibinfo {author} {\bibfnamefont {C.}~\bibnamefont {Brendel}},
  \bibinfo {author} {\bibfnamefont {M.}~\bibnamefont {Schmidt}}, \ and\
  \bibinfo {author} {\bibfnamefont {F.}~\bibnamefont {Marquardt}},\ }\href@noop
  {} {\bibfield  {journal} {\bibinfo  {journal} {Physical Review X}\ }\textbf
  {\bibinfo {volume} {5}},\ \bibinfo {pages} {031011} (\bibinfo {year}
  {2015})}\BibitemShut {NoStop}%
\bibitem [{\citenamefont {Silveirinha}(2018)}]{silveirinha2018proof}%
  \BibitemOpen
  \bibfield  {author} {\bibinfo {author} {\bibfnamefont {M.~G.}\ \bibnamefont
  {Silveirinha}},\ }\href@noop {} {\bibfield  {journal} {\bibinfo  {journal}
  {arXiv preprint arXiv:1804.02190}\ } (\bibinfo {year} {2018})}\BibitemShut
  {NoStop}%
\bibitem [{\citenamefont {Faure}\ and\ \citenamefont
  {Zhilinskii}(2000)}]{faure2000topological}%
  \BibitemOpen
  \bibfield  {author} {\bibinfo {author} {\bibfnamefont {F.}~\bibnamefont
  {Faure}}\ and\ \bibinfo {author} {\bibfnamefont {B.}~\bibnamefont
  {Zhilinskii}},\ }\href@noop {} {\bibfield  {journal} {\bibinfo  {journal}
  {Physical review letters}\ }\textbf {\bibinfo {volume} {85}},\ \bibinfo
  {pages} {960} (\bibinfo {year} {2000})}\BibitemShut {NoStop}%
\bibitem [{\citenamefont {Prodan}\ and\ \citenamefont
  {Prodan}(2009)}]{prodan2009topological}%
  \BibitemOpen
  \bibfield  {author} {\bibinfo {author} {\bibfnamefont {E.}~\bibnamefont
  {Prodan}}\ and\ \bibinfo {author} {\bibfnamefont {C.}~\bibnamefont
  {Prodan}},\ }\href@noop {} {\bibfield  {journal} {\bibinfo  {journal}
  {Physical review letters}\ }\textbf {\bibinfo {volume} {103}},\ \bibinfo
  {pages} {248101} (\bibinfo {year} {2009})}\BibitemShut {NoStop}%
\bibitem [{\citenamefont {Kane}\ and\ \citenamefont
  {Lubensky}(2014)}]{kane2014topological}%
  \BibitemOpen
  \bibfield  {author} {\bibinfo {author} {\bibfnamefont {C.}~\bibnamefont
  {Kane}}\ and\ \bibinfo {author} {\bibfnamefont {T.}~\bibnamefont
  {Lubensky}},\ }\href@noop {} {\bibfield  {journal} {\bibinfo  {journal}
  {Nature Physics}\ }\textbf {\bibinfo {volume} {10}},\ \bibinfo {pages} {39}
  (\bibinfo {year} {2014})}\BibitemShut {NoStop}%
\bibitem [{\citenamefont {S{\"u}sstrunk}\ and\ \citenamefont
  {Huber}(2015)}]{susstrunk2015observation}%
  \BibitemOpen
  \bibfield  {author} {\bibinfo {author} {\bibfnamefont {R.}~\bibnamefont
  {S{\"u}sstrunk}}\ and\ \bibinfo {author} {\bibfnamefont {S.~D.}\ \bibnamefont
  {Huber}},\ }\href@noop {} {\bibfield  {journal} {\bibinfo  {journal}
  {Science}\ }\textbf {\bibinfo {volume} {349}},\ \bibinfo {pages} {47}
  (\bibinfo {year} {2015})}\BibitemShut {NoStop}%
\bibitem [{\citenamefont {Yang}\ \emph {et~al.}(2015)\citenamefont {Yang},
  \citenamefont {Gao}, \citenamefont {Shi}, \citenamefont {Lin}, \citenamefont
  {Gao}, \citenamefont {Chong},\ and\ \citenamefont
  {Zhang}}]{yang2015topological}%
  \BibitemOpen
  \bibfield  {author} {\bibinfo {author} {\bibfnamefont {Z.}~\bibnamefont
  {Yang}}, \bibinfo {author} {\bibfnamefont {F.}~\bibnamefont {Gao}}, \bibinfo
  {author} {\bibfnamefont {X.}~\bibnamefont {Shi}}, \bibinfo {author}
  {\bibfnamefont {X.}~\bibnamefont {Lin}}, \bibinfo {author} {\bibfnamefont
  {Z.}~\bibnamefont {Gao}}, \bibinfo {author} {\bibfnamefont {Y.}~\bibnamefont
  {Chong}}, \ and\ \bibinfo {author} {\bibfnamefont {B.}~\bibnamefont
  {Zhang}},\ }\href@noop {} {\bibfield  {journal} {\bibinfo  {journal}
  {Physical review letters}\ }\textbf {\bibinfo {volume} {114}},\ \bibinfo
  {pages} {114301} (\bibinfo {year} {2015})}\BibitemShut {NoStop}%
\bibitem [{\citenamefont {Fleury}\ \emph {et~al.}(2016)\citenamefont {Fleury},
  \citenamefont {Khanikaev},\ and\ \citenamefont {Alu}}]{fleury2016floquet}%
  \BibitemOpen
  \bibfield  {author} {\bibinfo {author} {\bibfnamefont {R.}~\bibnamefont
  {Fleury}}, \bibinfo {author} {\bibfnamefont {A.~B.}\ \bibnamefont
  {Khanikaev}}, \ and\ \bibinfo {author} {\bibfnamefont {A.}~\bibnamefont
  {Alu}},\ }\href@noop {} {\bibfield  {journal} {\bibinfo  {journal} {Nature
  communications}\ }\textbf {\bibinfo {volume} {7}},\ \bibinfo {pages} {11744}
  (\bibinfo {year} {2016})}\BibitemShut {NoStop}%
\bibitem [{\citenamefont {Peri}\ \emph {et~al.}(2018)\citenamefont {Peri},
  \citenamefont {Serra-Garcia}, \citenamefont {Ilan},\ and\ \citenamefont
  {Huber}}]{peri2018axial}%
  \BibitemOpen
  \bibfield  {author} {\bibinfo {author} {\bibfnamefont {V.}~\bibnamefont
  {Peri}}, \bibinfo {author} {\bibfnamefont {M.}~\bibnamefont {Serra-Garcia}},
  \bibinfo {author} {\bibfnamefont {R.}~\bibnamefont {Ilan}}, \ and\ \bibinfo
  {author} {\bibfnamefont {S.~D.}\ \bibnamefont {Huber}},\ }\href@noop {}
  {\bibfield  {journal} {\bibinfo  {journal} {arXiv preprint arXiv:1806.09628}\
  } (\bibinfo {year} {2018})}\BibitemShut {NoStop}%
\bibitem [{\citenamefont {Delplace}\ \emph {et~al.}(2017)\citenamefont
  {Delplace}, \citenamefont {Marston},\ and\ \citenamefont
  {Venaille}}]{delplace2017topological}%
  \BibitemOpen
  \bibfield  {author} {\bibinfo {author} {\bibfnamefont {P.}~\bibnamefont
  {Delplace}}, \bibinfo {author} {\bibfnamefont {J.}~\bibnamefont {Marston}}, \
  and\ \bibinfo {author} {\bibfnamefont {A.}~\bibnamefont {Venaille}},\
  }\href@noop {} {\bibfield  {journal} {\bibinfo  {journal} {Science}\ ,\
  \bibinfo {pages} {eaan8819}} (\bibinfo {year} {2017})}\BibitemShut {NoStop}%
\bibitem [{\citenamefont {Perrot}\ \emph {et~al.}(2018)\citenamefont {Perrot},
  \citenamefont {Delplace},\ and\ \citenamefont
  {Venaille}}]{perrot2018topological}%
  \BibitemOpen
  \bibfield  {author} {\bibinfo {author} {\bibfnamefont {M.}~\bibnamefont
  {Perrot}}, \bibinfo {author} {\bibfnamefont {P.}~\bibnamefont {Delplace}}, \
  and\ \bibinfo {author} {\bibfnamefont {A.}~\bibnamefont {Venaille}},\
  }\href@noop {} {\bibfield  {journal} {\bibinfo  {journal} {arXiv preprint
  arXiv:1810.03328}\ } (\bibinfo {year} {2018})}\BibitemShut {NoStop}%
\bibitem [{\citenamefont {Jin}\ \emph {et~al.}(2016)\citenamefont {Jin},
  \citenamefont {Lu}, \citenamefont {Wang}, \citenamefont {Fang}, \citenamefont
  {Joannopoulos}, \citenamefont {Solja{\v{c}}i{\'c}}, \citenamefont {Fu},\ and\
  \citenamefont {Fang}}]{jin2016topological}%
  \BibitemOpen
  \bibfield  {author} {\bibinfo {author} {\bibfnamefont {D.}~\bibnamefont
  {Jin}}, \bibinfo {author} {\bibfnamefont {L.}~\bibnamefont {Lu}}, \bibinfo
  {author} {\bibfnamefont {Z.}~\bibnamefont {Wang}}, \bibinfo {author}
  {\bibfnamefont {C.}~\bibnamefont {Fang}}, \bibinfo {author} {\bibfnamefont
  {J.~D.}\ \bibnamefont {Joannopoulos}}, \bibinfo {author} {\bibfnamefont
  {M.}~\bibnamefont {Solja{\v{c}}i{\'c}}}, \bibinfo {author} {\bibfnamefont
  {L.}~\bibnamefont {Fu}}, \ and\ \bibinfo {author} {\bibfnamefont {N.~X.}\
  \bibnamefont {Fang}},\ }\href@noop {} {\bibfield  {journal} {\bibinfo
  {journal} {Nature communications}\ }\textbf {\bibinfo {volume} {7}},\
  \bibinfo {pages} {13486} (\bibinfo {year} {2016})}\BibitemShut {NoStop}%
\bibitem [{\citenamefont {Gao}\ \emph {et~al.}(2016)\citenamefont {Gao},
  \citenamefont {Yang}, \citenamefont {Lawrence}, \citenamefont {Fang},
  \citenamefont {B{\'e}ri},\ and\ \citenamefont {Zhang}}]{gao2016photonic}%
  \BibitemOpen
  \bibfield  {author} {\bibinfo {author} {\bibfnamefont {W.}~\bibnamefont
  {Gao}}, \bibinfo {author} {\bibfnamefont {B.}~\bibnamefont {Yang}}, \bibinfo
  {author} {\bibfnamefont {M.}~\bibnamefont {Lawrence}}, \bibinfo {author}
  {\bibfnamefont {F.}~\bibnamefont {Fang}}, \bibinfo {author} {\bibfnamefont
  {B.}~\bibnamefont {B{\'e}ri}}, \ and\ \bibinfo {author} {\bibfnamefont
  {S.}~\bibnamefont {Zhang}},\ }\href@noop {} {\bibfield  {journal} {\bibinfo
  {journal} {Nature communications}\ }\textbf {\bibinfo {volume} {7}},\
  \bibinfo {pages} {12435} (\bibinfo {year} {2016})}\BibitemShut {NoStop}%
\bibitem [{\citenamefont {Jin}\ \emph {et~al.}(2018)\citenamefont {Jin},
  \citenamefont {Xia}, \citenamefont {Christensen}, \citenamefont {Wang},
  \citenamefont {Fong}, \citenamefont {Freeman}, \citenamefont {Gardner},
  \citenamefont {Fallahi}, \citenamefont {Hu}, \citenamefont {Wang} \emph
  {et~al.}}]{jin2018magnetically}%
  \BibitemOpen
  \bibfield  {author} {\bibinfo {author} {\bibfnamefont {D.}~\bibnamefont
  {Jin}}, \bibinfo {author} {\bibfnamefont {Y.}~\bibnamefont {Xia}}, \bibinfo
  {author} {\bibfnamefont {T.}~\bibnamefont {Christensen}}, \bibinfo {author}
  {\bibfnamefont {S.}~\bibnamefont {Wang}}, \bibinfo {author} {\bibfnamefont
  {K.~Y.}\ \bibnamefont {Fong}}, \bibinfo {author} {\bibfnamefont
  {M.}~\bibnamefont {Freeman}}, \bibinfo {author} {\bibfnamefont {G.~C.}\
  \bibnamefont {Gardner}}, \bibinfo {author} {\bibfnamefont {S.}~\bibnamefont
  {Fallahi}}, \bibinfo {author} {\bibfnamefont {Q.}~\bibnamefont {Hu}},
  \bibinfo {author} {\bibfnamefont {Y.}~\bibnamefont {Wang}},  \emph {et~al.},\
  }\href@noop {} {\bibfield  {journal} {\bibinfo  {journal} {arXiv preprint
  arXiv:1803.02913}\ } (\bibinfo {year} {2018})}\BibitemShut {NoStop}%
\bibitem [{\citenamefont {Silveirinha}(2016)}]{silveirinha2016bulk}%
  \BibitemOpen
  \bibfield  {author} {\bibinfo {author} {\bibfnamefont {M.~G.}\ \bibnamefont
  {Silveirinha}},\ }\href@noop {} {\bibfield  {journal} {\bibinfo  {journal}
  {Physical Review B}\ }\textbf {\bibinfo {volume} {94}},\ \bibinfo {pages}
  {205105} (\bibinfo {year} {2016})}\BibitemShut {NoStop}%
\bibitem [{\citenamefont {Shankar}\ \emph {et~al.}(2017)\citenamefont
  {Shankar}, \citenamefont {Bowick},\ and\ \citenamefont
  {Marchetti}}]{shankar2017topological}%
  \BibitemOpen
  \bibfield  {author} {\bibinfo {author} {\bibfnamefont {S.}~\bibnamefont
  {Shankar}}, \bibinfo {author} {\bibfnamefont {M.~J.}\ \bibnamefont {Bowick}},
  \ and\ \bibinfo {author} {\bibfnamefont {M.~C.}\ \bibnamefont {Marchetti}},\
  }\href@noop {} {\bibfield  {journal} {\bibinfo  {journal} {Physical Review
  X}\ }\textbf {\bibinfo {volume} {7}},\ \bibinfo {pages} {031039} (\bibinfo
  {year} {2017})}\BibitemShut {NoStop}%
\bibitem [{\citenamefont {Souslov}\ \emph {et~al.}(2017)\citenamefont
  {Souslov}, \citenamefont {van Zuiden}, \citenamefont {Bartolo},\ and\
  \citenamefont {Vitelli}}]{souslov2017topological}%
  \BibitemOpen
  \bibfield  {author} {\bibinfo {author} {\bibfnamefont {A.}~\bibnamefont
  {Souslov}}, \bibinfo {author} {\bibfnamefont {B.~C.}\ \bibnamefont {van
  Zuiden}}, \bibinfo {author} {\bibfnamefont {D.}~\bibnamefont {Bartolo}}, \
  and\ \bibinfo {author} {\bibfnamefont {V.}~\bibnamefont {Vitelli}},\
  }\href@noop {} {\bibfield  {journal} {\bibinfo  {journal} {Nature Physics}\
  }\textbf {\bibinfo {volume} {13}},\ \bibinfo {pages} {1091} (\bibinfo {year}
  {2017})}\BibitemShut {NoStop}%
\bibitem [{\citenamefont {Souslov}\ \emph {et~al.}(2018)\citenamefont
  {Souslov}, \citenamefont {Dasbiswas}, \citenamefont {Vaikuntanathan},\ and\
  \citenamefont {Vitelli}}]{souslov2018topological}%
  \BibitemOpen
  \bibfield  {author} {\bibinfo {author} {\bibfnamefont {A.}~\bibnamefont
  {Souslov}}, \bibinfo {author} {\bibfnamefont {K.}~\bibnamefont {Dasbiswas}},
  \bibinfo {author} {\bibfnamefont {S.}~\bibnamefont {Vaikuntanathan}}, \ and\
  \bibinfo {author} {\bibfnamefont {V.}~\bibnamefont {Vitelli}},\ }\href@noop
  {} {\bibfield  {journal} {\bibinfo  {journal} {arXiv preprint
  arXiv:1802.09649}\ } (\bibinfo {year} {2018})}\BibitemShut {NoStop}%
\bibitem [{\citenamefont {Tauber}\ \emph {et~al.}(2018)\citenamefont {Tauber},
  \citenamefont {Delplace},\ and\ \citenamefont {Venaille}}]{tauber2018odd}%
  \BibitemOpen
  \bibfield  {author} {\bibinfo {author} {\bibfnamefont {C.}~\bibnamefont
  {Tauber}}, \bibinfo {author} {\bibfnamefont {P.}~\bibnamefont {Delplace}}, \
  and\ \bibinfo {author} {\bibfnamefont {A.}~\bibnamefont {Venaille}},\
  }\href@noop {} {\bibfield  {journal} {\bibinfo  {journal} {arXiv preprint
  arXiv:1812.05488}\ } (\bibinfo {year} {2018})}\BibitemShut {NoStop}%
\bibitem [{\citenamefont {Van~Mechelen}\ and\ \citenamefont
  {Jacob}(2018)}]{van2018photon}%
  \BibitemOpen
  \bibfield  {author} {\bibinfo {author} {\bibfnamefont {T.}~\bibnamefont
  {Van~Mechelen}}\ and\ \bibinfo {author} {\bibfnamefont {Z.}~\bibnamefont
  {Jacob}},\ }\href@noop {} {\bibfield  {journal} {\bibinfo  {journal} {arXiv
  preprint arXiv:1806.01395}\ } (\bibinfo {year} {2018})}\BibitemShut {NoStop}%
\bibitem [{\citenamefont {Iga}(1995)}]{iga1995transition}%
  \BibitemOpen
  \bibfield  {author} {\bibinfo {author} {\bibfnamefont {K.}~\bibnamefont
  {Iga}},\ }\href@noop {} {\bibfield  {journal} {\bibinfo  {journal} {Journal
  of Fluid Mechanics}\ }\textbf {\bibinfo {volume} {294}},\ \bibinfo {pages}
  {367} (\bibinfo {year} {1995})}\BibitemShut {NoStop}%
\bibitem [{\citenamefont {Li}\ \emph {et~al.}(2010)\citenamefont {Li},
  \citenamefont {Morpurgo}, \citenamefont {B{\"u}ttiker},\ and\ \citenamefont
  {Martin}}]{li2010marginality}%
  \BibitemOpen
  \bibfield  {author} {\bibinfo {author} {\bibfnamefont {J.}~\bibnamefont
  {Li}}, \bibinfo {author} {\bibfnamefont {A.~F.}\ \bibnamefont {Morpurgo}},
  \bibinfo {author} {\bibfnamefont {M.}~\bibnamefont {B{\"u}ttiker}}, \ and\
  \bibinfo {author} {\bibfnamefont {I.}~\bibnamefont {Martin}},\ }\href@noop {}
  {\bibfield  {journal} {\bibinfo  {journal} {Physical Review B}\ }\textbf
  {\bibinfo {volume} {82}},\ \bibinfo {pages} {245404} (\bibinfo {year}
  {2010})}\BibitemShut {NoStop}%
\bibitem [{\citenamefont {Iga}(2001)}]{iga2001transition}%
  \BibitemOpen
  \bibfield  {author} {\bibinfo {author} {\bibfnamefont {K.}~\bibnamefont
  {Iga}},\ }\href@noop {} {\bibfield  {journal} {\bibinfo  {journal} {Fluid
  Dynamics Research}\ }\textbf {\bibinfo {volume} {28}},\ \bibinfo {pages}
  {465} (\bibinfo {year} {2001})}\BibitemShut {NoStop}%
\bibitem [{\citenamefont {Meidan}\ \emph {et~al.}(2011)\citenamefont {Meidan},
  \citenamefont {Micklitz},\ and\ \citenamefont
  {Brouwer}}]{meidan2011topological}%
  \BibitemOpen
  \bibfield  {author} {\bibinfo {author} {\bibfnamefont {D.}~\bibnamefont
  {Meidan}}, \bibinfo {author} {\bibfnamefont {T.}~\bibnamefont {Micklitz}}, \
  and\ \bibinfo {author} {\bibfnamefont {P.~W.}\ \bibnamefont {Brouwer}},\
  }\href@noop {} {\bibfield  {journal} {\bibinfo  {journal} {Physical Review
  B}\ }\textbf {\bibinfo {volume} {84}},\ \bibinfo {pages} {195410} (\bibinfo
  {year} {2011})}\BibitemShut {NoStop}%
\bibitem [{\citenamefont {Fulga}\ \emph {et~al.}(2011)\citenamefont {Fulga},
  \citenamefont {Hassler}, \citenamefont {Akhmerov},\ and\ \citenamefont
  {Beenakker}}]{fulga2011scatt}%
  \BibitemOpen
  \bibfield  {author} {\bibinfo {author} {\bibfnamefont {I.~C.}\ \bibnamefont
  {Fulga}}, \bibinfo {author} {\bibfnamefont {F.}~\bibnamefont {Hassler}},
  \bibinfo {author} {\bibfnamefont {A.~R.}\ \bibnamefont {Akhmerov}}, \ and\
  \bibinfo {author} {\bibfnamefont {C.~W.~J.}\ \bibnamefont {Beenakker}},\
  }\href {\doibase 10.1103/PhysRevB.83.155429} {\bibfield  {journal} {\bibinfo
  {journal} {Phys. Rev. B}\ }\textbf {\bibinfo {volume} {83}},\ \bibinfo
  {pages} {155429} (\bibinfo {year} {2011})}\BibitemShut {NoStop}%
\bibitem [{\citenamefont {Fulga}\ \emph {et~al.}(2012)\citenamefont {Fulga},
  \citenamefont {Hassler},\ and\ \citenamefont
  {Akhmerov}}]{fulga2012scattering}%
  \BibitemOpen
  \bibfield  {author} {\bibinfo {author} {\bibfnamefont {I.~C.}\ \bibnamefont
  {Fulga}}, \bibinfo {author} {\bibfnamefont {F.}~\bibnamefont {Hassler}}, \
  and\ \bibinfo {author} {\bibfnamefont {A.~R.}\ \bibnamefont {Akhmerov}},\
  }\href@noop {} {\bibfield  {journal} {\bibinfo  {journal} {Physical Review
  B}\ }\textbf {\bibinfo {volume} {85}},\ \bibinfo {pages} {165409} (\bibinfo
  {year} {2012})}\BibitemShut {NoStop}%
\bibitem [{\citenamefont {Hu}\ \emph {et~al.}(2015)\citenamefont {Hu},
  \citenamefont {Pillay}, \citenamefont {Wu}, \citenamefont {Pasek},
  \citenamefont {Shum},\ and\ \citenamefont {Chong}}]{hu2015measurement}%
  \BibitemOpen
  \bibfield  {author} {\bibinfo {author} {\bibfnamefont {W.}~\bibnamefont
  {Hu}}, \bibinfo {author} {\bibfnamefont {J.~C.}\ \bibnamefont {Pillay}},
  \bibinfo {author} {\bibfnamefont {K.}~\bibnamefont {Wu}}, \bibinfo {author}
  {\bibfnamefont {M.}~\bibnamefont {Pasek}}, \bibinfo {author} {\bibfnamefont
  {P.~P.}\ \bibnamefont {Shum}}, \ and\ \bibinfo {author} {\bibfnamefont
  {Y.}~\bibnamefont {Chong}},\ }\href@noop {} {\bibfield  {journal} {\bibinfo
  {journal} {Physical Review X}\ }\textbf {\bibinfo {volume} {5}},\ \bibinfo
  {pages} {011012} (\bibinfo {year} {2015})}\BibitemShut {NoStop}%
\bibitem [{\citenamefont {Bal}(2017)}]{bal2017topological}%
  \BibitemOpen
  \bibfield  {author} {\bibinfo {author} {\bibfnamefont {G.}~\bibnamefont
  {Bal}},\ }\href@noop {} {\bibfield  {journal} {\bibinfo  {journal} {arXiv
  preprint arXiv:1709.00605}\ } (\bibinfo {year} {2017})}\BibitemShut {NoStop}%
\bibitem [{\citenamefont {Fefferman}\ \emph {et~al.}(2016)\citenamefont
  {Fefferman}, \citenamefont {Lee-Thorp},\ and\ \citenamefont
  {Weinstein}}]{fefferman2016edge}%
  \BibitemOpen
  \bibfield  {author} {\bibinfo {author} {\bibfnamefont {C.~L.}\ \bibnamefont
  {Fefferman}}, \bibinfo {author} {\bibfnamefont {J.~P.}\ \bibnamefont
  {Lee-Thorp}}, \ and\ \bibinfo {author} {\bibfnamefont {M.~I.}\ \bibnamefont
  {Weinstein}},\ }\href@noop {} {\bibfield  {journal} {\bibinfo  {journal}
  {Annals of PDE}\ }\textbf {\bibinfo {volume} {2}},\ \bibinfo {pages} {12}
  (\bibinfo {year} {2016})}\BibitemShut {NoStop}%
\bibitem [{\citenamefont {Bal}(2018)}]{bal2018continuous}%
  \BibitemOpen
  \bibfield  {author} {\bibinfo {author} {\bibfnamefont {G.}~\bibnamefont
  {Bal}},\ }\href@noop {} {\bibfield  {journal} {\bibinfo  {journal} {arXiv
  preprint arXiv:1808.07908}\ } (\bibinfo {year} {2018})}\BibitemShut {NoStop}%
\bibitem [{\citenamefont {Faure}(2019)}]{faure2019manifestation}%
  \BibitemOpen
  \bibfield  {author} {\bibinfo {author} {\bibfnamefont {F.}~\bibnamefont
  {Faure}},\ }\href@noop {} {\bibfield  {journal} {\bibinfo  {journal} {arXiv
  preprint arXiv:1901.10592}\ } (\bibinfo {year} {2019})}\BibitemShut {NoStop}%
\bibitem [{\citenamefont {Drouot}(2019)}]{drouot2019bulk}%
  \BibitemOpen
  \bibfield  {author} {\bibinfo {author} {\bibfnamefont {A.}~\bibnamefont
  {Drouot}},\ }\href@noop {} {\bibfield  {journal} {\bibinfo  {journal} {arXiv
  preprint arXiv:1901.06281}\ } (\bibinfo {year} {2019})}\BibitemShut {NoStop}%
\bibitem [{Note1()}]{Note1}%
  \BibitemOpen
  \bibinfo {note} {Edge numbers $n_\protect \mathrm {top}$ and $n_\protect
  \mathrm {bottom}$ are defined up to a global sign, depending on the
  orientation of the boundary, but the relative sign between them in their
  definition persists anyway.}\BibitemShut {Stop}%
\bibitem [{\citenamefont {Thomson}(1880)}]{thomson1880}%
  \BibitemOpen
  \bibfield  {author} {\bibinfo {author} {\bibfnamefont {W.}~\bibnamefont
  {Thomson}},\ }\href@noop {} {\bibfield  {journal} {\bibinfo  {journal}
  {Proceedings of the Royal Society of Edinburgh}\ }\textbf {\bibinfo {volume}
  {10}},\ \bibinfo {pages} {92} (\bibinfo {year} {1880})}\BibitemShut {NoStop}%
\bibitem [{\citenamefont {Faure}\ and\ \citenamefont
  {Zhilinskii}(2001)}]{faure2001topological}%
  \BibitemOpen
  \bibfield  {author} {\bibinfo {author} {\bibfnamefont {F.}~\bibnamefont
  {Faure}}\ and\ \bibinfo {author} {\bibfnamefont {B.}~\bibnamefont
  {Zhilinskii}},\ }\href@noop {} {\bibfield  {journal} {\bibinfo  {journal}
  {Letters in Mathematical Physics}\ }\textbf {\bibinfo {volume} {55}},\
  \bibinfo {pages} {219} (\bibinfo {year} {2001})}\BibitemShut {NoStop}%
\bibitem [{\citenamefont {Fukui}\ \emph {et~al.}(2012)\citenamefont {Fukui},
  \citenamefont {Shiozaki}, \citenamefont {Fujiwara},\ and\ \citenamefont
  {Fujimoto}}]{fukui2012bulk}%
  \BibitemOpen
  \bibfield  {author} {\bibinfo {author} {\bibfnamefont {T.}~\bibnamefont
  {Fukui}}, \bibinfo {author} {\bibfnamefont {K.}~\bibnamefont {Shiozaki}},
  \bibinfo {author} {\bibfnamefont {T.}~\bibnamefont {Fujiwara}}, \ and\
  \bibinfo {author} {\bibfnamefont {S.}~\bibnamefont {Fujimoto}},\ }\href@noop
  {} {\bibfield  {journal} {\bibinfo  {journal} {Journal of the Physical
  Society of Japan}\ }\textbf {\bibinfo {volume} {81}},\ \bibinfo {pages}
  {114602} (\bibinfo {year} {2012})}\BibitemShut {NoStop}%
\end{thebibliography}%
\end{document}